\documentclass[a4paper,fleqn,usenatbib]{mnras}
\usepackage[T1]{fontenc}
\usepackage{ae,aecompl}
\usepackage{graphicx}
\usepackage{footnote}
\usepackage[fleqn]{amsmath}
\usepackage[varg]{txfonts}
\usepackage{multirow}
\usepackage{subfig}

\title[Constraining the duty cycle of transient LMXBs through simulations]
{Constraining the duty cycle of transient low-mass X-ray binaries through simulations}

\author[D. Carbone \& R. Wijnands]
{
D.~Carbone$^{1,2}$\thanks{E-mail: dario.carbone86@gmail.com}, R.~Wijnands$^{2}$\\
$^{1}$ Department of Physics and Astronomy, Texas Tech University, Box 1051, Lubbock, TX 79409-1051, USA \\
$^{2}$ Anton Pannekoek Institute for Astronomy, University of Amsterdam, Postbus 94249, 1090 GE Amsterdam, The Netherlands
}

\date{Accepted June 11, 2019}

\begin{document}
\maketitle

\begin{abstract}
We performed simulations of a large number of so-called very faint X-ray transient sources from surveys obtained using the
X-ray telescope aboard the Neil Gehrels \emph{Swift} Observatory on two Galactic globular clusters, and the Galactic Center.
We calculated the ratio between the duty cycle we input in our simulations and the one we measure after the simulations.
We found that fluctuations in outburst duration and recurrence times affect our estimation of the duty cycle more than
non detected outbursts. This biases our measures to overestimate the simulated duty cycle of sources.
Moreover, we determined that compact surveys are necessary to detect outbursts with short duration because they could fall in
gaps between observations, if such gaps are longer than their duration.
On the other hand, long surveys are necessary to detect sources with low duty cycle because the smallest duty cycle a survey
can observe is given by the ratio between the shortest outburst duration and the total length of the survey.
If one has a limited amount of observing time, these two effects are competing, and a compromise is required which is set by the
goals of the proposed survey.
We have also performed simulations with several artificial survey strategies in order to evaluate the optimal observing
campaign aimed at detecting transients as well as at having the most accurate estimates of the duty cycle.
As expected, the best campaign would be a regular and dense monitoring that extends for a very long period.
The closest real example of such a dataset is the monitoring of the Galactic Centre.
\end{abstract}


\section{Introduction}
X-ray binaries are constituted of a compact object (a neutron star or a black hole) accreting mass from a companion. If the companion
is a relatively low mass star (typically of the order $\lesssim$\,1\,M$_{\odot}$) then those systems are called low-mass X-ray binaries
(LMXBs). Most LMXBs are transient sources: they are usually in a very faint quiescent state but occasionally they show bright X-ray
outbursts \cite[for typical outburst light curves see][]{Chen1997,Yan2015}. \citet{Wijnands2006} classified transient LMXBs according
to their peak X-ray luminosity (in the energy range 2-10 keV) during outbursts into multiple classes\footnote{It is important to stress that this classification
is mostly based on observed phenomenology and has no immediate connections to any astrophysical property of these systems. In addition,
some systems have shown outbursts with a large variety in their peak luminosities and therefore would fall in different classes depending on which of
their outbursts is used in the classification.}. The bright to very bright transients have peak outburst luminosities of 10$^{37-39}$\,erg\,s$^{-1}$.
Due to their brightness those systems are readily discovered and have been intensively studied over the last 4 decades. Therefore, we have a good
understanding of their behavior. 

Faint and very faint X-ray transients (VFXTs) have peak X-ray luminosities of 10$^{36-37}$\,erg\,s$^{-1}$ and 10$^{34-36}$\,erg\,s$^{-1}$,
respectively. Their faintness makes outbursts of those systems significantly more difficult to discover compared to those of the brighter transients
because the resulting fluxes are typically low to very low and often below the sensitivity of X-ray all-sky monitors (ASMs) orbiting the Earth.
This problem is of course most severe for the VFXTs and such systems are typically discovered serendipitously when sensitive X-ray satellites
(e.g., \emph{Swift}, \emph{Chandra}, \emph{XMM-Newton}) are pointed at certain sky positions in the Galaxy \citep[for early examples of the
discovery of VFXTs see, e.g.,][]{Hands2004, Muno2005, Porquet2005} or at Galactic globular clusters \citep[e.g.,][]{Heinke2009, Heinke2010}.
To increase the probability of finding new VFXTs and detecting more outbursts of the known systems, several optimised observing campaigns
have been performed both for the Galactic centre region \citep[][]{Wijnands2006, Degenaar2010, Degenaar2012,Degenaar2013,Degenaar2015}
as well as targeting several globular clusters \cite[][]{Altamirano2011ATel, Wijnands2012ATel, Linares2016ATel}.

The current leading model to explain the mechanism behind the outbursts of transient LMXBs is the disk instability model \citep[DIM; for an extensive
review on the DIM see][]{Lasota2001}. Between the outbursts, the material supplied by the companion star is stored in the relatively cool disk
surrounding the compact object. This eventually leads to a thermal instability which results in the increase of the viscosity and therefore the accretion
rate onto the compact object resulting in a X-ray outburst. Currently it is not yet clear if the DIM can also explain the outburst of the VFXTs, but
\citet[][]{Hameury2016} argued that indeed the DIM could explain those systems if they are so-called ultra-compact X-ray binaries which are systems
with a very small orbital period of $\lesssim$90 minutes and have hydrogen poor accretion disks.

The duty cycle (DC) of a transient source is expressed as the ratio between the time spent in outburst and the time interval between the start
of two consecutive outbursts (thus the recurrence time of the outbursts).
Both the outburst and the recurrence times are very important ingredients
in the DIM \cite[see details in][]{Lasota2001}, so determining accurate  DCs is crucial to constrain and test the DIM.
Determining accurate DCs is important for several other reasons as well. Using the DC and 
the averaged mass accretion rate during outbursts (which can be obtained from the
averaged source luminosity in outburst in combination with the source distance),
we can obtain the averaged (over the outburst and quiescent period;
assuming no accretion at all takes place in quiescence) mass accretion rate ($\langle\dot{M}_{\textrm A}\rangle$). Since the historical outburst behavior
of X-ray transients is only known for years to at most several decades, this average can only be calculated from the observational data for a similar
time span. Therefore, we have to assume that this estimate is representative for the mass accretion rate over time scales as long as $>$10$^{4-6}$
years. For such long periods, $\langle\dot{M}_{\textrm A}\rangle$ can be assumed to be equal to the mass transfer rate
$\langle\dot{M}_{\textrm T}\rangle$ from the companion star (assuming conservative mass transfer; thus any mass loss due to outflows,
like jets or winds, is assumed to be negligible). This $\langle\dot{M}_{\textrm T}\rangle$ is an important parameter in X-ray binary evolution
models  \citep[see, e.g., the review by][]{Tauris2006} as well as population synthesis models of LMXBs
\citep[e.g.,][and references therein]{Haaften2013, Haaften2015}.

In addition, determining an accurate $\langle\dot{M}_{\textrm A}\rangle$ (and thus obtaining accurate DCs) is also important in several studies
involving neutron star physics. For example knowing $\langle\dot{M}_{\textrm A}\rangle$ is crucial in understanding how neutron stars are heated
due to accretion of matter and cool when the accretion has halted \citep[e.g.,][]{Brown1998, Colpi2001, Yakovlev2004, Wijnands2008, Heinke2009_1,
Wijnands2013, Wijnands2017, Ootesetal2019}. Determining an accurate $\langle\dot{M}_{\textrm A}\rangle$ is also very important in understanding
the magnetic field evolution in accreting neutron stars, both
for the long term evolution \citep[][]{Taam1986, Romani1990, Geppert1994} as well as for possible short term screening of the magnetic field by the
accreted matter, and why some neutron-star LMXBs exhibit millisecond pulsations and others do not \citep[e.g.,][]{Cumming2001, Galloway2006,
Wijnands2008, Patruno2012_1, Patruno2012}. Finally, knowing $\langle\dot{M}_{\textrm A}\rangle$ accurately is important to determine the physical
reason why the spin distribution of neutron stars appears to have a cut-off at about $\sim$730 Hz \citep[e.g.,][]{Chakrabarty2003, Patruno2012_2,
Papitto2014}.

We have relatively good constraints on the DC of the bright and very bright transients because their outbursts are detectable with ASMs, and many
of them have exhibited multiple outbursts. Typically the DC of those transients is in the range 0.01-0.1, with an average value of $\sim$0.03
\citep[see, e.g., Figure 10 in][]{Yan2015}. We note that those DCs are determined over a time span of about 15 years, and it is not necessarily true
that they are representative for the long-term (i.e., evolutionary time scales) DCs of those systems. This short time span introduces a
detection bias for systems that have large DCs because those systems are more likely to have multiple detected outbursts, and therefore that
their DCs can be constrained. 

Due to the difficulties in detecting outbursts of faint X-ray transients and VFXTs, many of those outbursts are missed, significantly hampering
the determination of the DCs for those systems (again this is most severe for the VFXTs). Based on 4 years of Swift/XRT monitoring of the Galactic
centre, \citet[][]{Degenaar2010} found a large range of DCs for the detected VFXTs in their survey, ranging from DCs of only a few percent
\citep[see also][]{Degenaar2015} to DCs above 50\%\footnote{Several VFXTs with such high DCs are known \citep[][]{DelSanto2007, Degenaar2010,
Arnason2015} and are commonly referred to as quasi-persistent transients \citep[e.g., see][who was the first, to our knowledge, who called these
transients this way]{Remillard1999} which are transient LMXB that have very long, up to decades, outbursts. Examples of quasi-persistent transients
can be found in all luminosity classes defined in this paper.}. For this reason, the DCs of VFXTs appear very similar to that of the brighter transients
\citep[see][]{Yan2015}.
Constraining accurate DCs for the VFXTs might also help to differentiate the different potential sub-types of VFXTs \citep[see][]{Wijnands2006} from
each other, although investigating this is beyond the scope of our current paper.

The observing campaigns in which outbursts of VFXTs are typically detected are very infrequent, often not regularly spaced in time, and often
with very large time span between observations. It is thus likely that the calculated DCs (or the upper limits on the DCs) for many VFXTs have large
uncertainties. The aim of this paper is to investigate how the accuracy of the DCs of VFXTs is affected by the properties of the observing campaigns.

Several existing observing campaigns can be used for our study. For example, the Galactic centre has been very frequently monitored
\citep[nearly once every day for $>$10 years now;][]{Degenaar2010, GCmagnetardiscovery_Degenaar2013, Degenaar2015} using the X-ray
telescope (XRT) aboard \emph{Swift}. Several VFXTs have indeed been detected, in some case even with multiple outbursts
\citep[see, e.g., the summary given in][]{Degenaar2015}. In addition, several observing campaign  \citep[e.g., using \emph{RXTE} or
\emph{Swift}/XRT;][]{Altamirano2011ATel, Wijnands2012ATel, Linares2016ATel} on Galactic Globular cluster systems have been preformed
to find X-ray transients, since a sizable number
of those clusters are expected to host multiple transient LMXBs (e.g., \citealt[][]{Heinke2003}; see Table 5 of \citealt[][]{Bahramian2014} for a list of
active, both persistent as well as transient, LMXBs in globular clusters). Therefore, pointed observations using sensitive X-ray instrumentation of
those clusters will allow to monitor several systems at once increasing the likelihood that outbursts are discovered. Because of the sensitive of the
XRT in combination with the flexibility of \emph{Swift}, this satellite is currently most often used to obtain such pointings (as well as pointings at the
Galactic centre). For this reason, in our simulations we focus only on observing campaigns performed with \emph{Swift}/XRT.

Beside using the sampling of existing campaigns we will also investigate if we can determine what kind of observing strategy would constrain the
DCs best, while maximizing the detection of outbursts (thus allowing us to determine the optimal observing strategy to discover more VFXTs). We
will discuss the \emph{Swift}/XRT observing campaign we use in Section~\ref{sec:observations}; in Section~\ref{sec:methods}
we present the methods used to perform our simulations. We will then present the results of our investigation in Section~\ref{sec:results} and discuss
their implications and conclude in Section~\ref{sec:conclusion}.

\section{Observing Strategies}
\label{sec:observations}

Using \emph{Swift}/XRT\footnote{http://www.swift.ac.uk/swift\_portal/index.php}, a number of Globular clusters have been monitored frequently,
either to find more transients \citep[][]{Wijnands2012ATel, Linares2016ATel} or to monitor detected transients during their outbursts
\citep[e.g.,][]{DelSanto2014, Bahramian2014, Linares2014, Tetarenko2016}. Those campaigns are excellent input for our simulations because they
demonstrate directly what is possible using \emph{Swift}/XRT, and are therefore representative of the accuracy one can obtain on the DCs. From the
clusters monitored with \emph{Swift}/XRT we decided to use the observations from NGC~6388 (discussed in Section~\ref{sec:ngc6388}), and
Terzan~5 (discussed in Section~\ref{sec:terzan5}).

NGC~6388 was chosen because it only had a three month period during which it was monitored about once a week. The separation between
consecutive observations is smaller than the minimum outbursts duration we have assumed in our simulations (see Section~\ref{sec:methods}),
implying that it is very unlikely that a simulated outburst occurring during the input observing campaign would not be detected. On the other hand,
the time span of this survey is very limited, making it very difficult to recover transients with a low DC.
Terzan~5 was chosen because the observations cover a relatively long time span of 4.5 years, with episodes of dense sampling, but also with large
gaps between observations. These gaps range between a few days, and almost 2 years.  The presence of long gaps will likely cause many
outbursts to be non detected, allowing us to determine the effect of this on the accuracy of the observed DCs.

The observing campaign on these clusters are quite different, allowing us to investigate the effects of those different strategies on the obtained DCs.
We note that we use those campaigns only as sampling strategy; whether or not a source was detected (or even monitored) during those observations
is irrelevant for the purpose of our paper.
Table~\ref{tab:observations} summarises the observing dates that were used in the simulations for the surveys of NGC~6388 and Terzan~5,
highlighting the difference between the two.

The observing campaign of the Galactic Centre includes a total of 1682 observations (due to the very large number of observations we do not
include them in Table~\ref{tab:observations}), covering a period between October 2005 and November 2017 with gaps of about 3 months between
November and February due to Solar constraints. This is the best \emph{Swift}/XRT dataset for a single position in the sky because it has a long
baseline and a very dense coverage.

\section{Simulations}
\label{sec:methods}

We have used the simulation technique developed by \citet{Carbone2017}. We simulated 10$^5$ individual sources for each survey strategy
separately.  The light curves are modelled as a fast rise, exponential decay type of outburst, and are fully characterised by their start time, their
peak luminosity, and the time it takes for their luminosity to decay below 10$^{34}$~erg~s$^{-1}$ (from now on, duration). We are aware that
many observed outbursts do not have profiles as the one we have assumed  \citep[e.g.,][]{Chen1997, Yan2015}, but the exact shape of the light
curve has a negligible effect on the estimation of the DC\footnote{
This is true because the quantity that affects the {\textrm DC} the most is the amount of matter that is accreted during an outburst. In fact, this
would affect the duration of the following quiescence period, because it would take the source a different amount of time to replenish its accretion
disk. The amount of matter that is accreted is mostly determined by the peak luminosity and the duration of an outburst, rather than by the exact
light curve shape (as far as it is not radically different).
.}
However, the exact shape would strongly matter when converting the DC to other properties of the sources, such as
$\langle\dot{M}_{\textrm A}\rangle$. Calculating these properties and how they are affected by assuming different outbursts profiles is beyond
the scope of our current paper.

In our simulations, we assumed that outbursts from the same source will have similar peak luminosities, similar durations,
and that intervals between the outbursts will be similar as well. We therefore associated a single value for the peak luminosity and the
duration of outbursts, and for the DC to each source.
The peak luminosity of the outbursts has been simulated uniformly in logarithmic space between 10$^{34}$ and
10$^{36}$~erg~sec$^{-1}$. We did not simulate brighter outbursts because they would not require dedicated observations
using high sensitivity telescopes.
The outburst peak luminosity from the same source is usually very similar between different outbursts, although
variations are seen \citep[e.g.,][]{Yan2015}.
For this reason we allow for a variation of a factor of 2 in the peak luminosity of different outbursts of the same source.
The actual values and variability in the outburst peak luminosities are irrelevant for the current paper, but 
we included them in our code for completeness and for future works that will, e.g., calculate $\langle\dot{M}_{\textrm A}\rangle$ as well.
The outburst durations have been simulated uniformly in logarithmic space between 7 and 200 days
\citep[see, e.g.,][for typical outburst durations]{Yan2015}. Also in this case, we allowed it to vary up to a factor of two for
different outbursts of the same source.
Finally, each source has a simulated DC, randomly chosen between 0.0001 and 0.15.
The DC of a source is the ratio between the time spent by a source in outburst and the time interval between
the start of two consecutive outbursts:

\begin{equation}
\textrm{DC} = \frac{T_{{\textrm {outburst}},\,i}}{t_{{\textrm{start}},\,i+1} - t_{{\textrm{start}},\,i}} \: ,
\label{eq:dc}
\end{equation}

\noindent where $T_{\textrm{outburst}, i}$ and $t_{\textrm{start}, i}$ are the duration and the start time of the current outburst and 
$t_{\textrm{start}, i+1}$ is the start time of the following outburst of the same source.
We allowed the DC of a source to vary up to a factor of 2 in both directions between different outbursts of the same source.
In order to determine the start time of the first outburst we simulated for a source (with given DC and T$_{\textrm{outburst}}$),
we first calculated the earliest time at which such source could have started an outburst, assuming its next one happens after our observation
campaign is ongoing. In order to do so, we used Equation~\ref{eq:dc}, where t$_{\textrm{start}, i+1}$ is equal to the beginning of our observing
campaign, and solved for t$_{\textrm{start}, i}$.
The interval between this value and the start of our observations defines a whole cycle for our source, as shifting the start time of an
outburst between these extreme is equivalent to observing a source at different times throughout its cycle.
The start time of the first outburst we simulated for each source is therefore uniformly distributed between the two values we just referred to.
Using again Equation~\ref{eq:dc}, we calculate the start time the next outburst of the same source, and if this time is lower than the end
of the survey, the following outburst will be simulated in the same manner.
This way we produced a catalog with simulated sources, each of which is constituted by multiple outbursts, and has a certain
value of the DC.

We also have to input a survey strategy for which we used the previously mentioned observing campaigns
(see Section~\ref{sec:observations}).
The most important information needed from those campaigns is the start time of the observations. Our code also requires
to input the integration time and the sensitivity of each observation \citep[see][]{Carbone2017}.
The sensitivity of our observations is determined by our instrument, in this case \emph{Swift}/XRT.
For sources at the distance of the targets we chose to simulate (NGC~6388, Terzan~5, and the Galactic Center),
\emph{Swift}/XRT is sensitive enough to detect sources until they enter the quiescence period, i.e. when their luminosity drops 
below  10$^{34}$\,erg\,s$^{-1}$ \citep{Plotkin2013,Wijnands2015}.
We therefore define the limit for when a source is in outburst as to when its luminosity is above 10$^{34}$\,erg\,s$^{-1}$.

The integration time is less important because most of the outbursts have peak luminosity much larger than the sensitivity limit, but it is still
relevant because some of them might be detectable only during the decay phase, and their signal might be smeared out in the background noise
in long observations.

This list of simulated sources is then checked against the survey strategy to test if each of the outburst we simulated is detected or not.
This is done by checking if the outburst was active during each observation.
(i.e., if it was bright enough to be detected).
This way we produced a catalog of the
simulated sources that were detected at least once; for each source a different number of outbursts have been detected. After this, we calculated
the DC for each of the sources that were detected at least twice, so we can compare it to the input DC.
This calculation is performed as explained in Section~\ref{sec:results}.
We calculated a value of the DC for each pair of consecutive detected outbursts. The DC of a source for which many outbursts
are detected is calculated both as the maximum and as the average of the calculated DCs. We chose to calculate both because if an outburst
from one such source has not been detected, the average DC would be strongly biased towards lower values, whereas the maximum would
still be as close as possible to the real one. On the other hand, the maximum would be systematically biased towards larger values, not
representing the real value as well as the average, if the latter is not dramatically affected by missing outbursts. Our simulations only consider sources
that exhibit at least one outburst during the considered observing campaign because we are interested in testing how obtained DCs compare
to the input DCs, i.e., we want to test how accurate the measurements of DCs from real data are and in what way they could be biased.

\section{Results}
\label{sec:results}

We define the value of DC that was inputted in our simulations as {\textrm DC}$_{\textrm{sim}}$ (the simulated DC) and the value of DC that
is calculated from the catalog of the detected sources as {\textrm DC}$_{\textrm{obs}}$ (observed DC, although observed in this context means
determined from our simulated datasets and not obtained from actual observations). {\textrm DC}$_{\textrm{obs}}$ is calculated using
Equation~\ref{eq:dc}, where the numerator, indicating the duration of outburst $i$ (T$_{\textrm{outburst},\,i}$), is derived from the observations,
i.e. it is the time difference between the first and last detection of such outburst. This value is different from the simulated duration of that event
because an outburst might not be detected as soon as it starts, and might not be observed as it fades away completely. The denominator,
indicating the interval between the first detections of two consecutive detected outbursts $i$, and $i+1$, is also derived from the observations
(i.e., from our simulated datasets).
This value can be different from the simulated one both because an outburst might not be detected as soon as it starts, and because outbursts
might end up being non detected at all, strongly affecting this calculation (i.e., two consecutive \textit{detected} outbursts might not be
two consecutive \textit{simulated} outbursts).
As a consequence of how we modelled the DC and the duration of the outbursts, {\textrm DC}$_{\textrm{obs}}$ can be as high as 4 times
{\textrm DC}$_{\textrm{sim}}$ because both the numerator and the denominator of Equation~\ref{eq:dc} can vary by up to a factor of 2.
We repeated our simulations allowing the duration of the outbursts to vary by a factor of 3 instead of 2, and then the maximum
{\textrm DC}$_{\textrm{obs}}$ was indeed 6 times {\textrm DC}$_{\textrm{sim}}$ as expected. In future work we will include a probability
distribution on those quantities in our code and this could change the maximum ratio between {\textrm DC}$_{\textrm{obs}}$ and
{\textrm DC}$_{\textrm{sim}}$, however, we expect this will not change the main conclusions of our paper.

As previously mentioned, due to our simulation setup there are sources that are in outburst before the
campaign started and do not repeat while it is ongoing; those sources are excluded when calculating the probabilities discussed below.

\begin{figure*}
\centering
\includegraphics[scale=0.47,viewport = 0 0 520 410, clip]{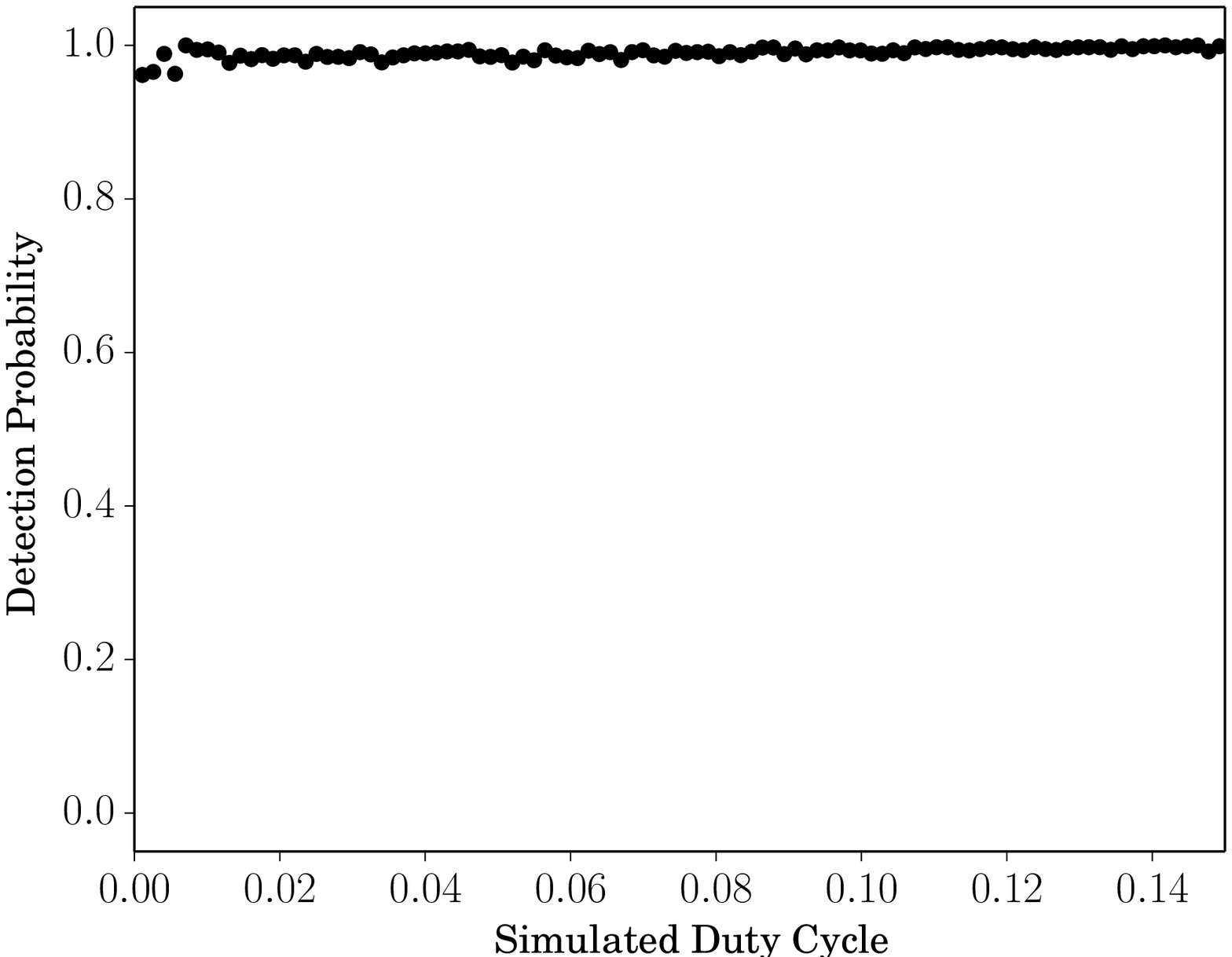}
\includegraphics[scale=0.47,viewport = 0 0 530 410, clip]{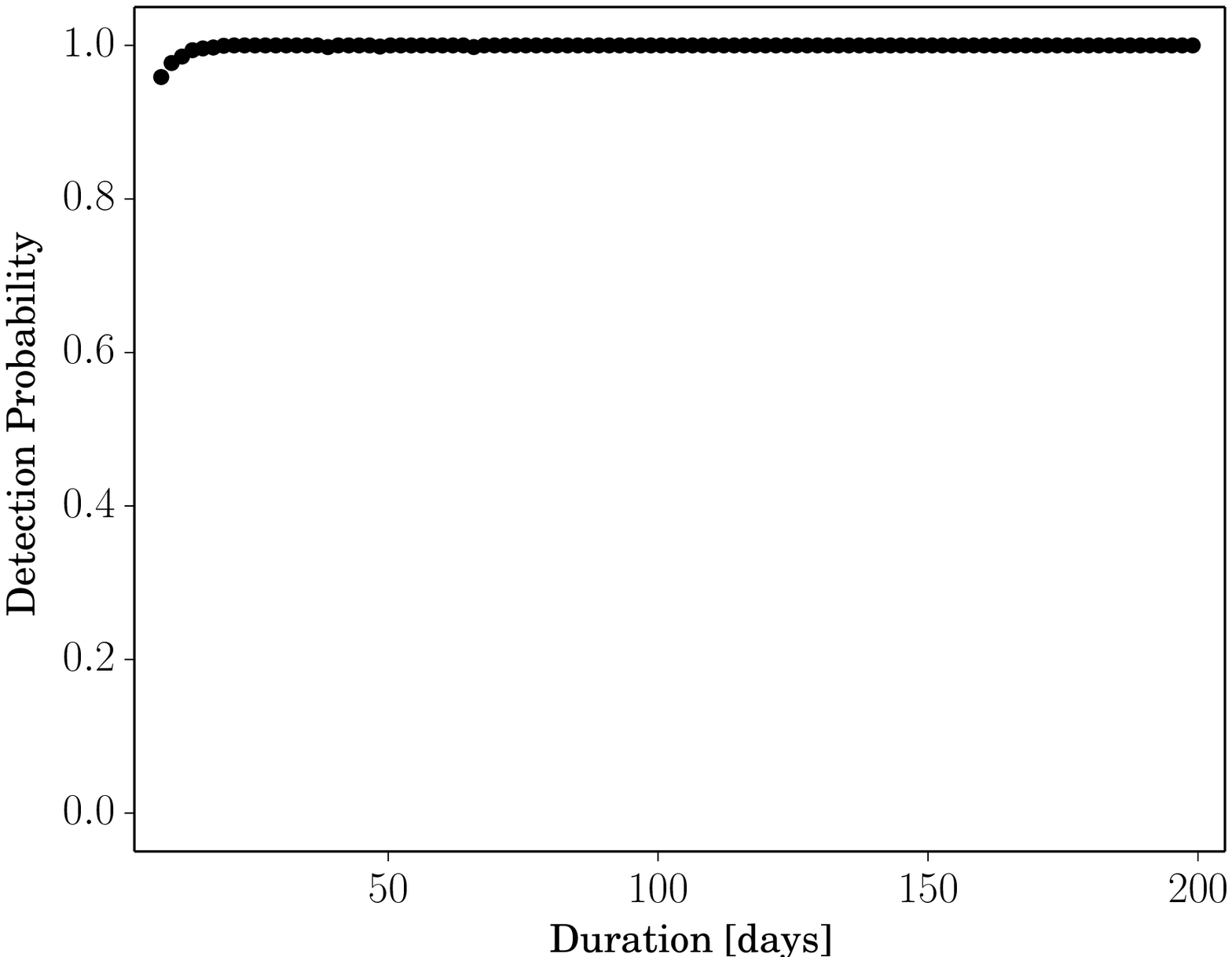}
\includegraphics[scale=0.46,viewport = 0 0 555 410, clip]{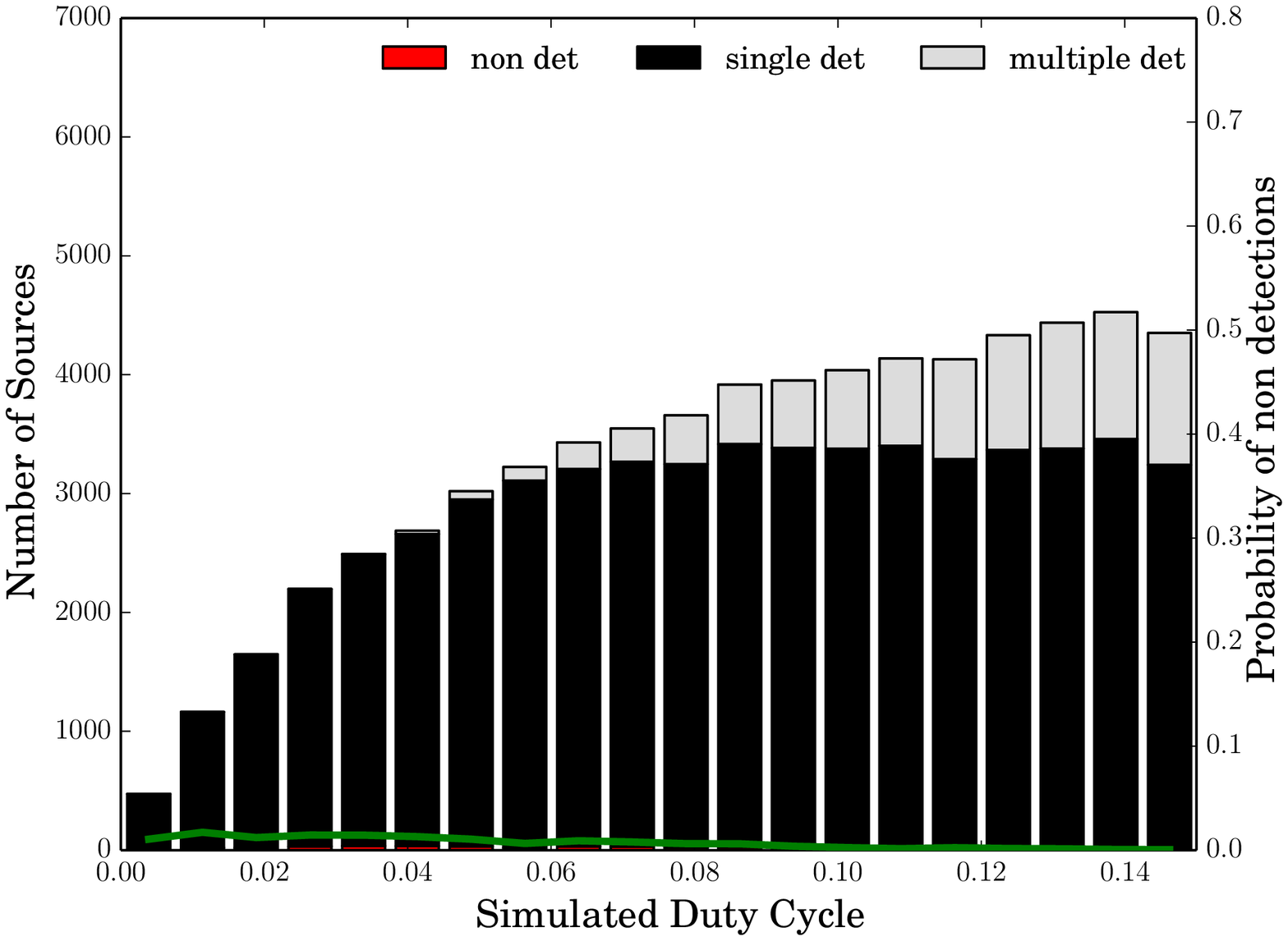}
\includegraphics[scale=0.47,viewport = 0 0 520 410, clip]{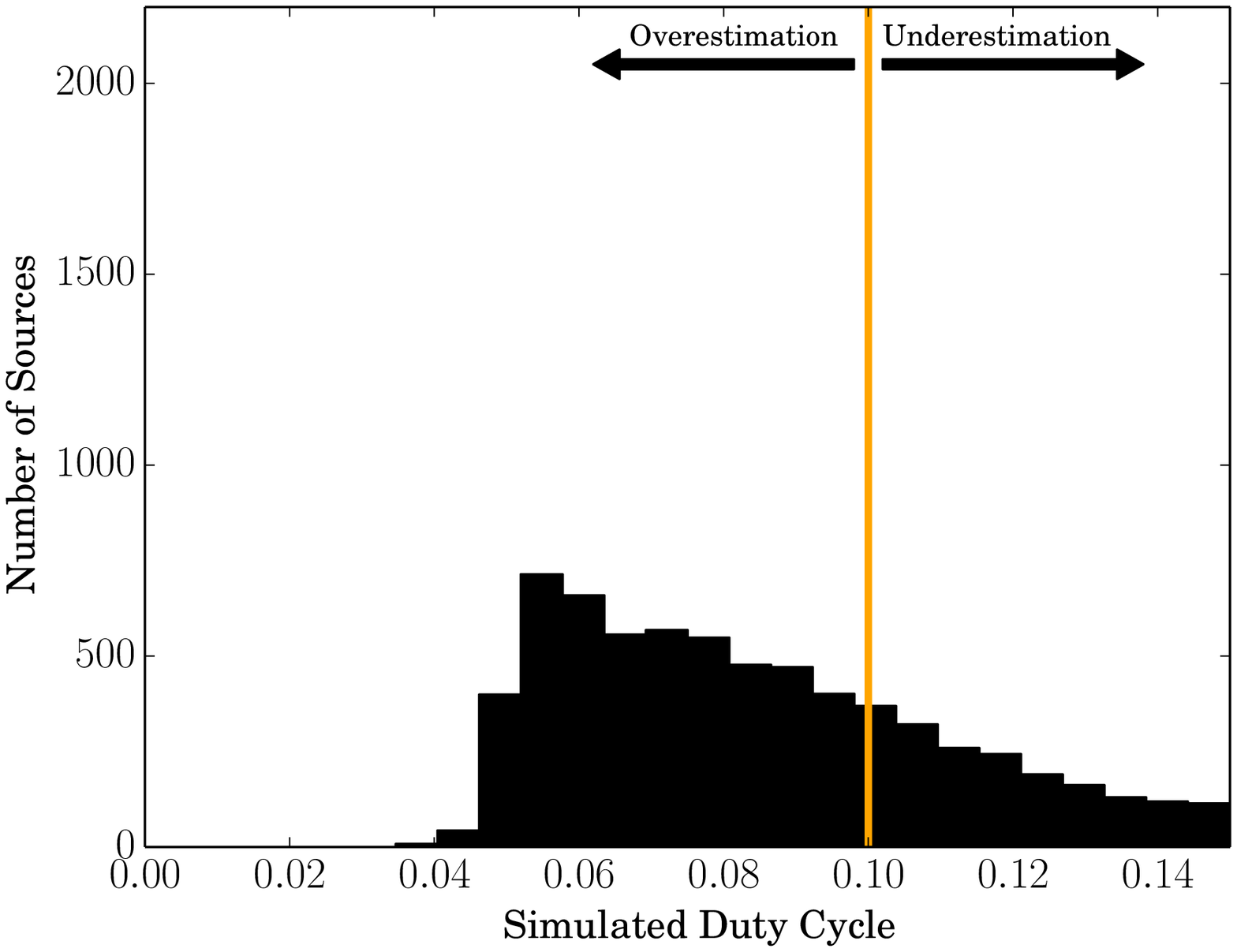}
\includegraphics[scale=0.435,viewport = 0 0 570 410, clip]{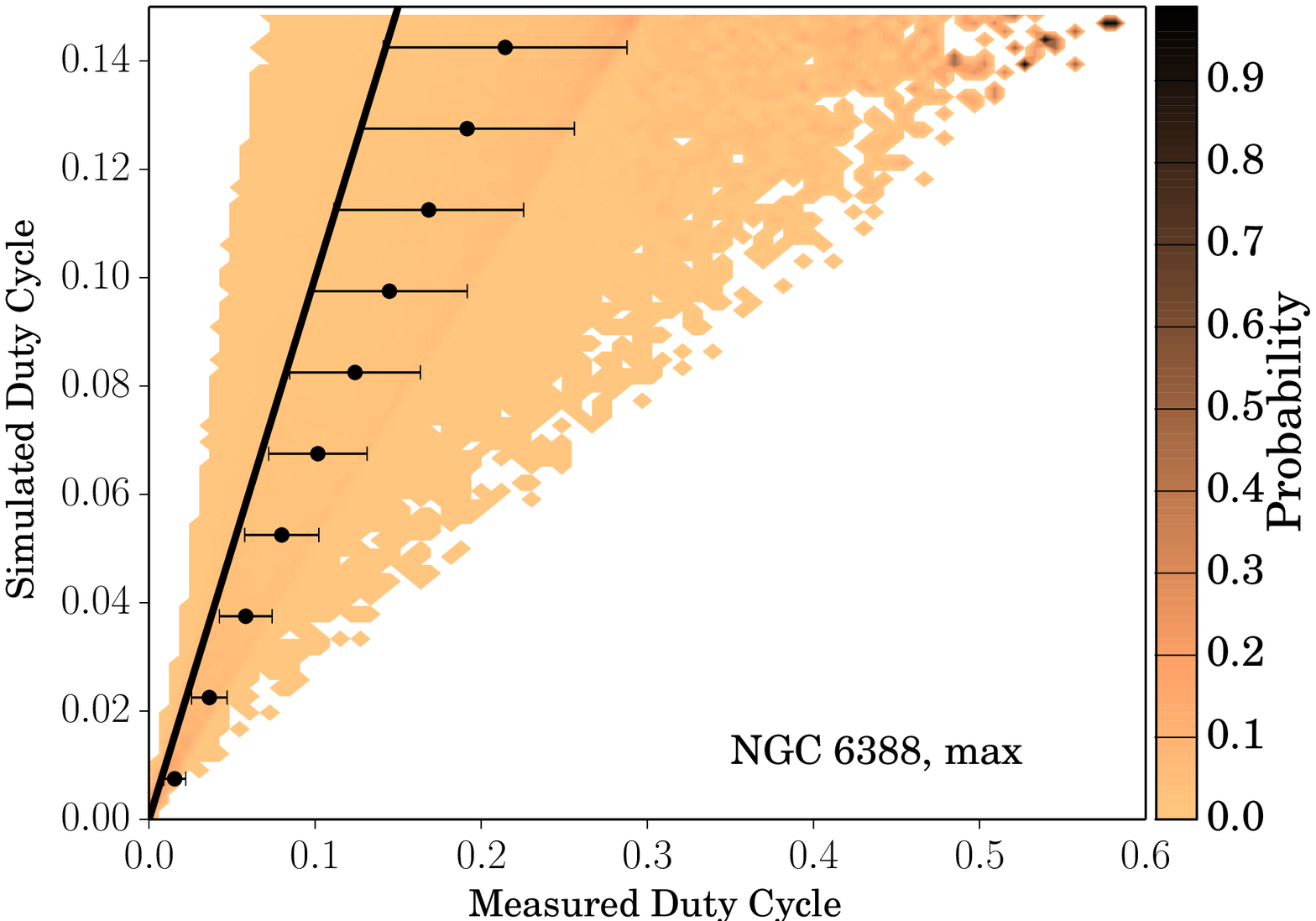}
\includegraphics[scale=0.435,viewport = 0 0 570 410, clip]{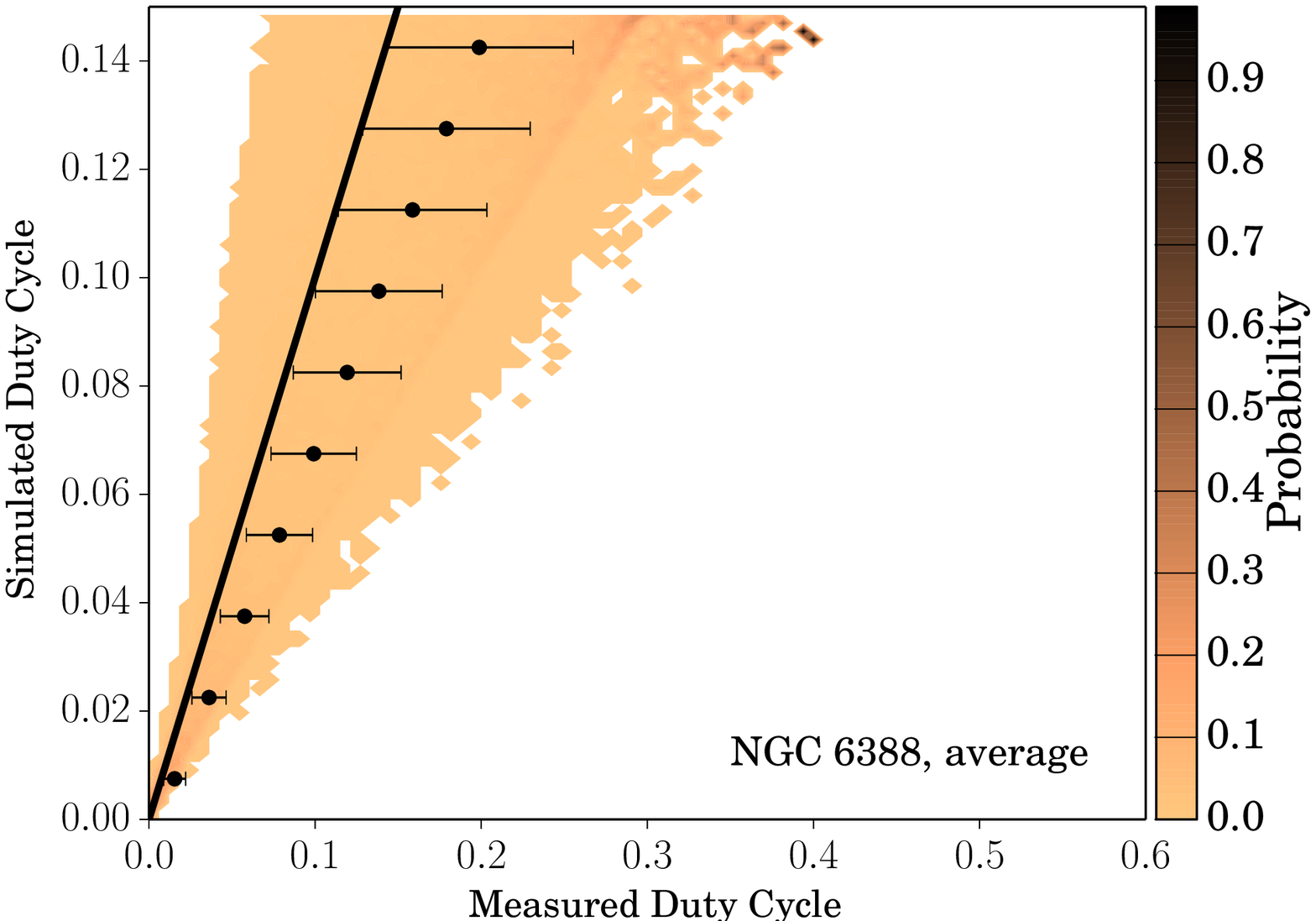}
\caption{
Results of our simulations using the observing campaign of NGC~6388 as input survey strategy. Only sources exhibiting at least one
outburst during the observing campaign are presented in this analysis. The top left panel shows the probability of detecting
a source as a function of its {\textrm DC}$_{\textrm{sim}}$; the top right panel shows the probability of detecting an outburst as a function
of its duration. The middle left panel represents a cumulative histogram of all simulated sources color coded based on the number of
outbursts that were detected. The green line in that plot (visible at the very bottom in this case), represents the probability that a source
would never be detected (i.e., all outbursts from that source would not be detected) as a function of its {\textrm DC}$_{\textrm{sim}}$.
The middle right panel represents a cross-cut at {\textrm DC}$_{\textrm{obs}}$\,=\,0.10 in the bottom right panel. It shows how accurate
our estimation is. The bottom panels show, colour coded, the probability that a source has a certain simulated DC
({\textrm DC}$_{\textrm{sim}}$) provided that the observed value ({\textrm DC}$_{\textrm{obs}}$) is another. The sum of the probability
is one in each vertical bin. The black dots in the bottom panels represent the average value of the simulated DC for different observed ones.
In the bottom left panel {\textrm DC}$_{\textrm{obs}}$ is calculated as the maximum from different outburst pairs, while in the bottom
right panel it is calculated as the average. The black line in both bottom panels represents the ideal one-to-one relation.
}
\label{fig:ngc6388}
\end{figure*}

\subsection{NGC~6388}
\label{sec:ngc6388}
For our first simulations, we used the \emph{Swift}/XRT observing campaign of NGC~6388 as our survey strategy.
The results of those simulations are shown in Figure~\ref{fig:ngc6388}.
The detection probability discussed here and for the rest of the manuscript is defined as the ratio between the number
of detected sources with a certain value of {\textrm DC}$_{\textrm{sim}}$ (or other variables) and the total number of simulated sources
with the same {\textrm DC}$_{\textrm{sim}}$ (or other variables). As mentioned earlier, only sources that exhibit at least one outburst during
the observing campaign are taken into account in this calculation. A source is detected if at least one outburst is detected.
We calculated the probability of detecting transient sources as a function of their {\textrm DC}$_{\textrm{sim}}$ 
(see top left panel of Figure~\ref{fig:ngc6388}) and found that it is very high ($\sim$\,1) for all the values of DC.
This is due to the fact that every outburst that occurs during the campaign is detected thanks to the compactness of the campaign.
Only very few outbursts are missed, those that have only a small portion happening during an observation.
We also observe that there is a scatter especially at low values of DC.
This is due to the fact that for such low values, only one outburst occurred during the observing campaign.
If this outburst was not detected, then the whole source was not, and therefore it does not show up in the plot, whereas for
higher values of DC, if one outburst was missed, another one from the same source could still have been detected, increasing
the probability that the source is detected.

We also calculated the probability of detecting sources as a function of the average duration of their outbursts. 
We note that almost all sources were detected, with a probability of detection of 1 for almost all durations.
Only very short outbursts are sometimes non detected. This is due to the fact that outbursts
shorter than 14 days may appear in only one observation and, for the shortest, their luminosity might have dropped below
10$^{34}$~erg~s$^{-1}$ at the time of the observation.
This is shown in the top right panel of Figure~\ref{fig:ngc6388}.

We then identified and selected sources that have multiple detected outbursts.
From the middle left panel of Figure~\ref{fig:ngc6388} we can see that almost all sources exhibiting an outburst during the observing campaign
have been detected at least one, but also that only a minority of sources have been detected multiple times. All the multiple detections belong
to sources with {\textrm DC}$_{\textrm{sim}}\,\geq\,0.04$. These effects (all sources are detected, but only a minority multiple times) are due to the very
dense coverage of the observations, and to the very short duration of the campaign respectively.
We have simulated an even number of sources per DC bin. The sources missing from the bins with low DC in this panel are the ones that do not
exhibit any outburst during the observing campaign and are therefore excluded from this analysis. This effect is not a surprise as sources with low DC
are rarer, and therefore if the first outburst we simulated happened before the campaign started, then the following one would happen only after the
same campaign was over, and as the observations of NGC~6388 lasted only for a few months, several sources suffered this effect.
This will be very different for different strategies as highlighted in the remainder of the manuscript.
The green line in the same plot represents the probability that a source would not be detected with this observing campaign, i.e.,
all of the outbursts from this source would not be detected.
In the case of NGC~6388 this probability is very small, lower than 2\% for all values of the duty cycle.
This quantity depends a lot on the observing strategy, and will be very different for different campaigns.

Finally, we compared {\textrm DC}$_{\textrm{obs}}$ with {\textrm DC}$_{\textrm{sim}}$, both when {\textrm DC}$_{\textrm{obs}}$ is calculated as
the maximum and when it is calculated as the average of the observed DC for each source.
The two cases are represented in the bottom left and bottom right panels of Figure~\ref{fig:ngc6388} respectively.
For different values of {\textrm DC}$_{\textrm{obs}}$ we calculated the probability that such value corresponds
to specific {\textrm DC}$_{\textrm{sim}}$, i.e., we considered all sources having a certain value of {\textrm DC}$_{\textrm{obs}}$
and compared it to their {\textrm DC}$_{\textrm{sim}}$.
The black line represents the ideal one-to-one relation.
We note that the two plots look similar because almost all sources have at most two detected outbursts,
meaning that the maximum and the average {\textrm DC}$_{\textrm{obs}}$ are the same.
For almost all values of {\textrm DC}$_{\textrm{obs}}$ it is possible to both overestimate and underestimate {\textrm DC}$_{\textrm{sim}}$.
In order to quantify this discrepancy, we show in black dots the average value of {\textrm DC}$_{\textrm{obs}}$ for each bin of
{\textrm DC}$_{\textrm{sim}}$, and we note that the average {\textrm DC}$_{\textrm{obs}}$ is systematically larger than {\textrm DC}$_{\textrm{sim}}$,
despite having a large scatter. This is more clearly visible in the middle right panel of Figure~\ref{fig:ngc6388}, where we performed a cross-cut at
{\textrm DC}$_{\textrm{obs}}$\,=\,0.10 in the bottom right panel (the one using the average {\textrm DC}$_{\textrm{obs}}$).
From this plot we see that there are more sources for which we overestimated their DC, rather underestimating it.
This is expected, as we can more easily detect outbursts from sources which have  more frequent outbursts than the average.
We remind that we introduced a factor of 2 scatter in the DC.

\begin{figure*}
\centering
\includegraphics[scale=0.47,viewport = 0 0 520 410, clip]{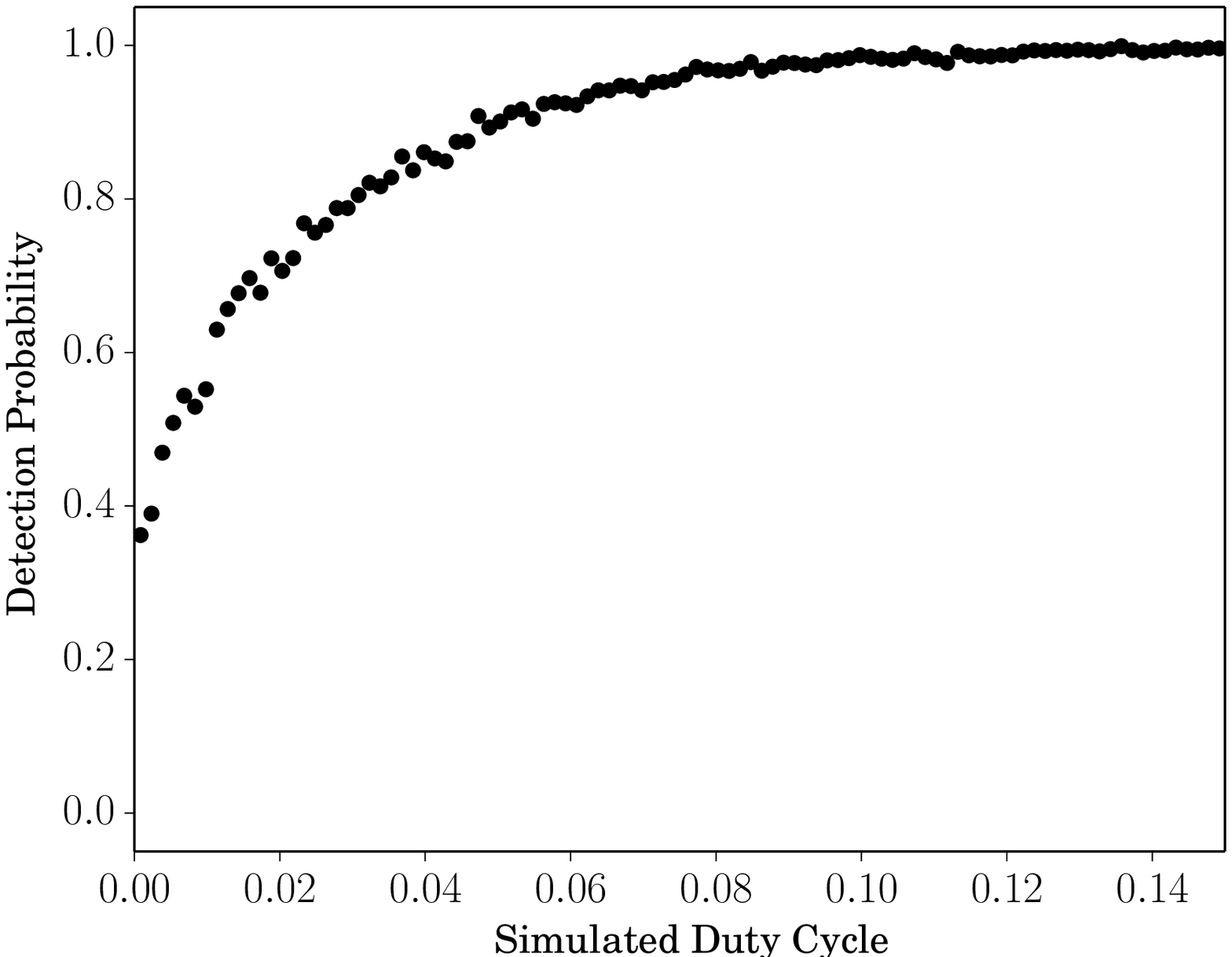}
\includegraphics[scale=0.47,viewport = 0 0 530 410, clip]{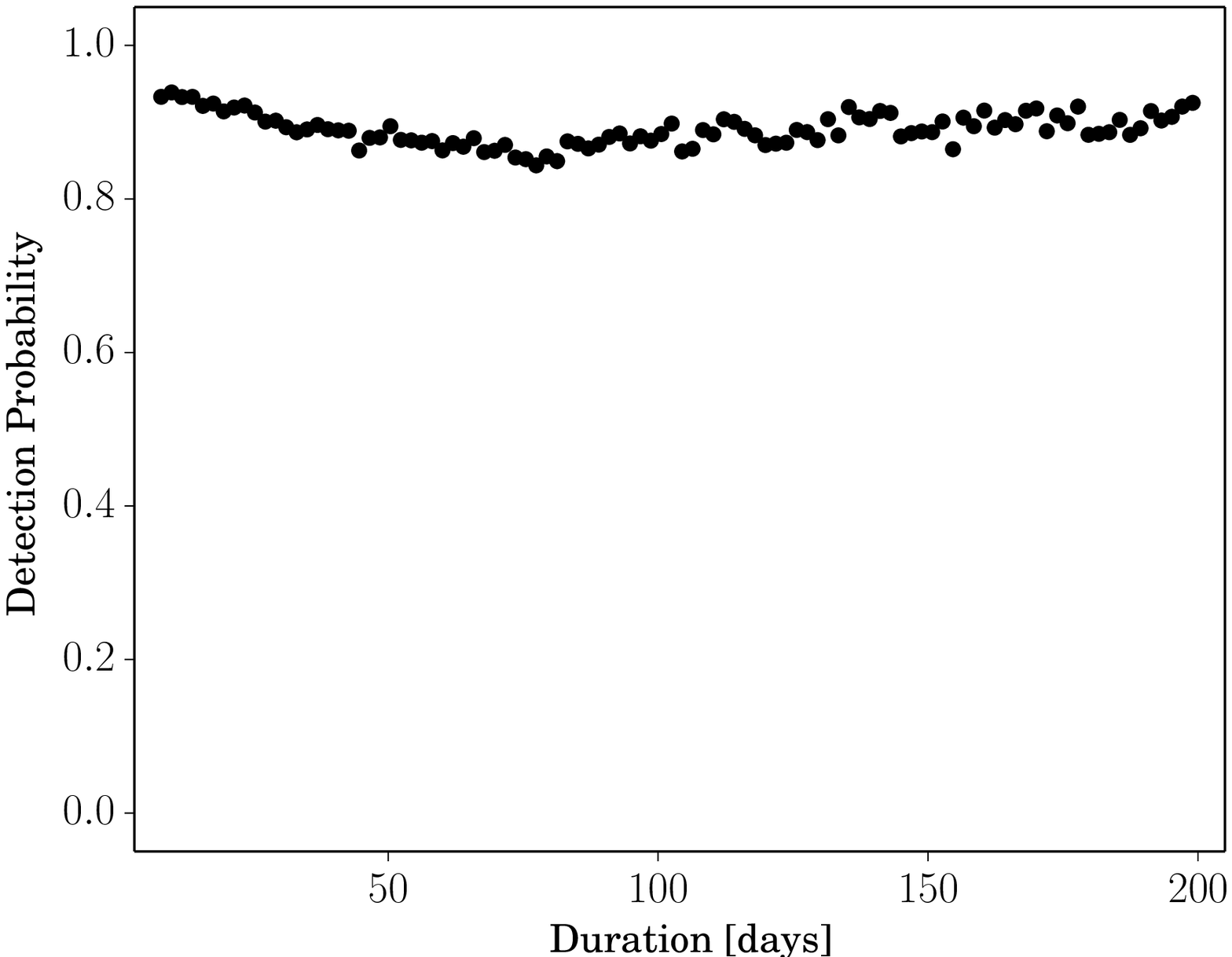}
\includegraphics[scale=0.46,viewport = 0 0 555 410, clip]{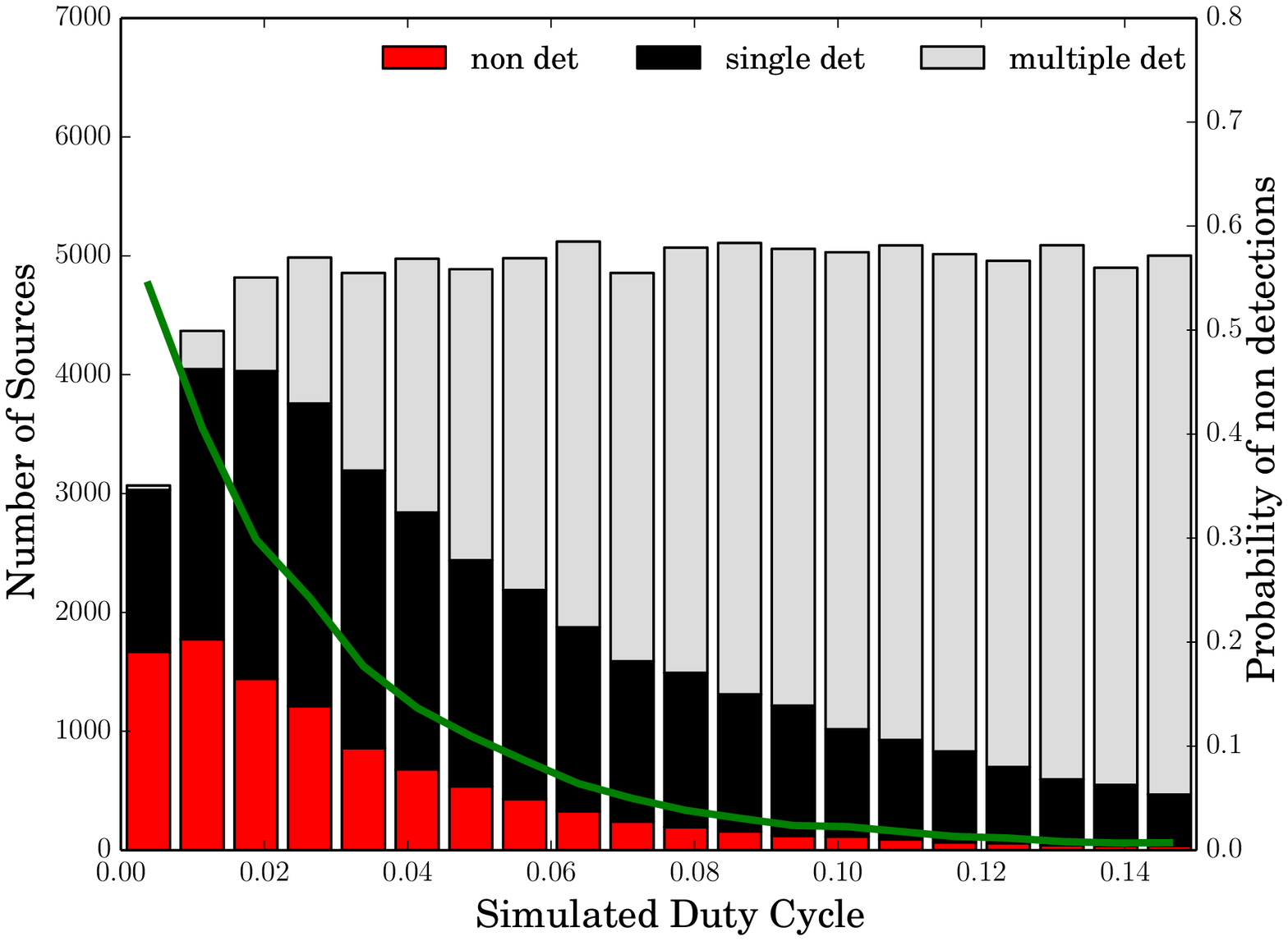}
\includegraphics[scale=0.47,viewport = 0 0 520 410, clip]{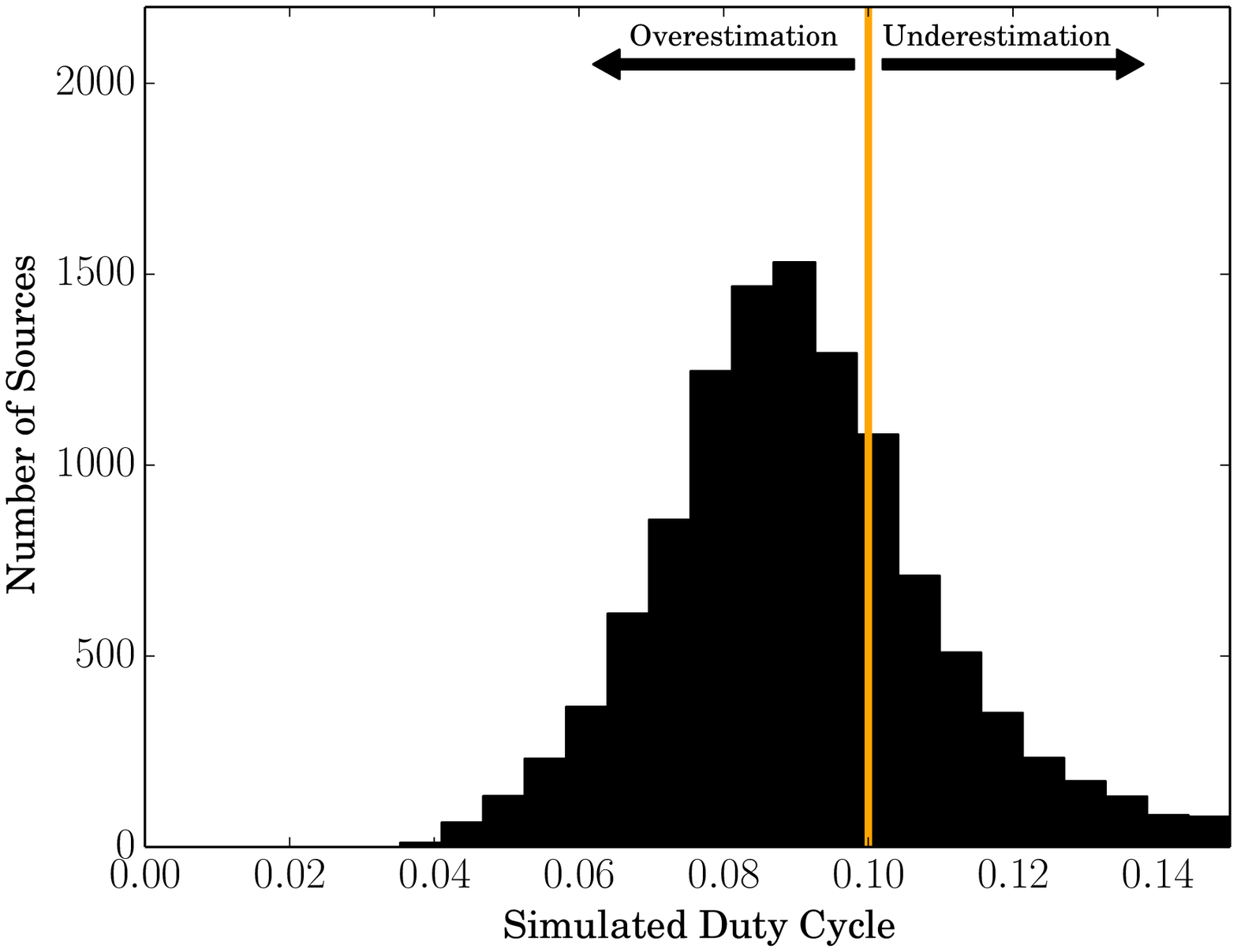}
\includegraphics[scale=0.435,viewport = 0 0 570 410, clip]{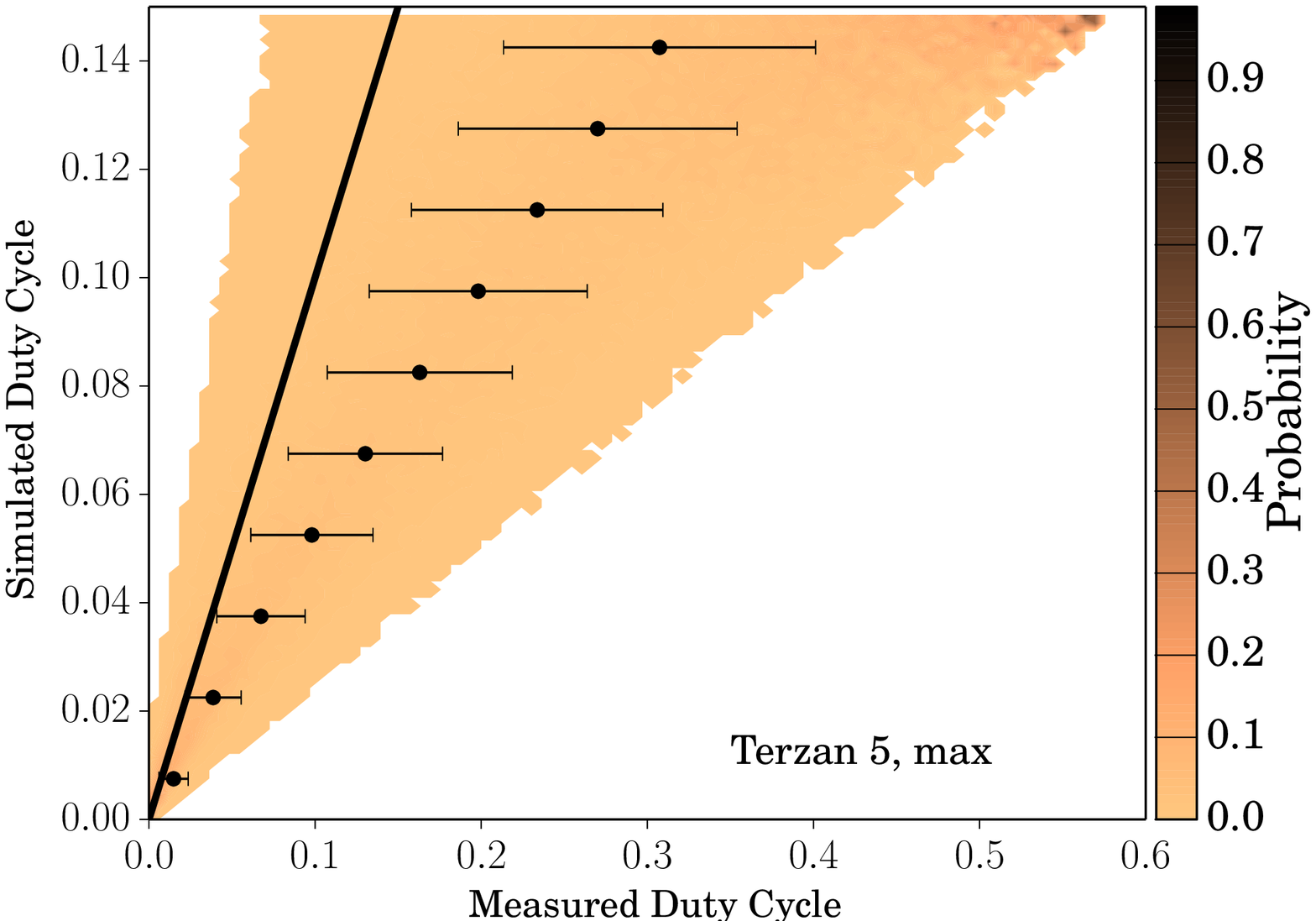}
\includegraphics[scale=0.435,viewport = 0 0 570 410, clip]{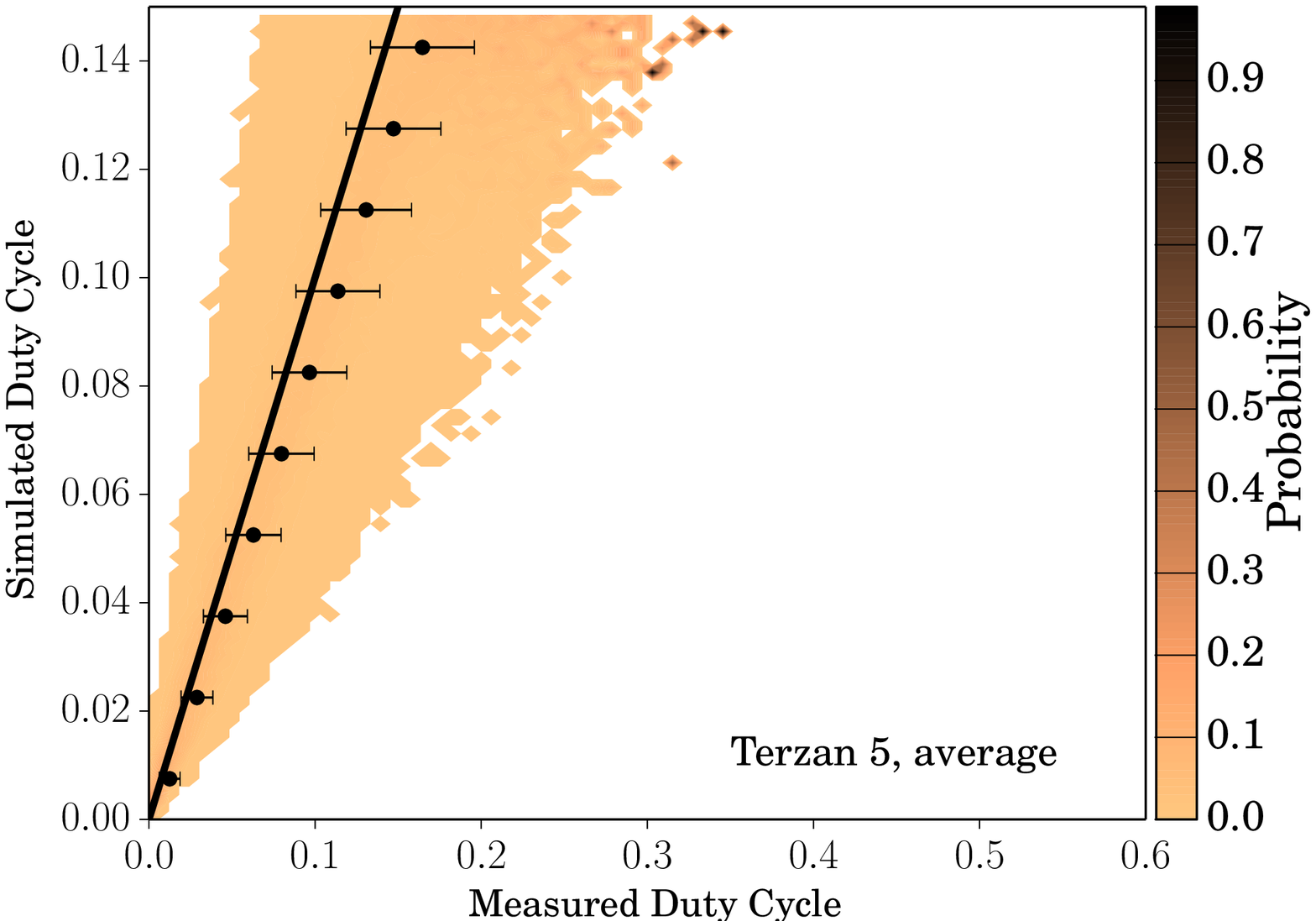}
\caption{
Results of our simulations using the observing campaign of Terzan~5 as input survey strategy.
The six panels are the same as in Figure~\ref{fig:ngc6388}.
}
\label{fig:terz5}
\end{figure*}

\subsection{Terzan~5}
\label{sec:terzan5}
We have performed the same analysis for the globular cluster Terzan~5, which has a rather different observing campaign
than NGC~6388 (see Section~\ref{sec:observations}). The outcome of our simulations is presented in Figure~\ref{fig:terz5}.
In the top left panel we can see that in this case the probability of detecting sources as a function of their {\textrm DC}$_{\textrm{sim}}$
goes from about 0.4 for very low values of {\textrm DC}$_{\textrm{sim}}$, reaching 1 for the highest values of {\textrm DC}$_{\textrm{sim}}$.
Similar to our simulations of NGC~6388, this is due to the fact that more frequently recurring sources (i.e. high DC)
exhibit more outbursts during the observing campaign, increasing the probability of detecting at least one.
The probability to detect simulated outbursts as a function of their duration is plotted in the top right panel of Figure~\ref{fig:terz5}.
Here the probability of detection remains above 0.8 for all values of the duration, but does not stay constant at 1 for long outbursts.
This plot is very different than the one observed for NGC~6388 and is due to the presence of long gaps (>\,200d) between consecutive
observations. Sources of all durations might go into outburst during a gap and such outburst, if too short, might not be detected.
If such outburst is the only one happening during the campaign the whole source will never be detected, whereas if 
the recurrence time is shorter, other outbursts could compensate for such eventuality.
This implies that most of the sources that are not detected have low {\textrm DC}$_{\textrm{sim}}$, as visible in the middle left panel.

On one hand, the presence of long gaps cause some sources to end up non detected, most of which have small {\textrm DC}$_{\textrm{sim}}$.
On the other hand, the long duration of the observing campaign allows for a large number of sources to have multiple detected outbursts.
In the case of Terzan~5 the probability that a source would not be detected reaches values above 50\% for sources with
very small duty cycles, and decreases monotonically, as already mentioned.

In the case of the observing campaign of Terzan~5, the two bottom panels are more diverse compared to the campaign of NGC~6388.
The spread of the colored area, as well as the error bars on the black dot are larger in the bottom left panel,
where {\textrm DC}$_{\textrm{obs}}$ is calculated as the maximum, compared to the bottom right panel
(where {\textrm DC}$_{\textrm{obs}}$ is calculated as the average).
Also for Terzan~5, {\textrm DC}$_{\textrm{obs}}$ can both be an overestimation or an underestimation of {\textrm DC}$_{\textrm{sim}}$,
but the overestimations are more common, as highlighted by the positions of the black dots, and by the middle
right panel in Figure \ref{fig:terz5}.

\begin{figure*}
\centering
\includegraphics[scale=0.47,viewport = 0 0 520 410, clip]{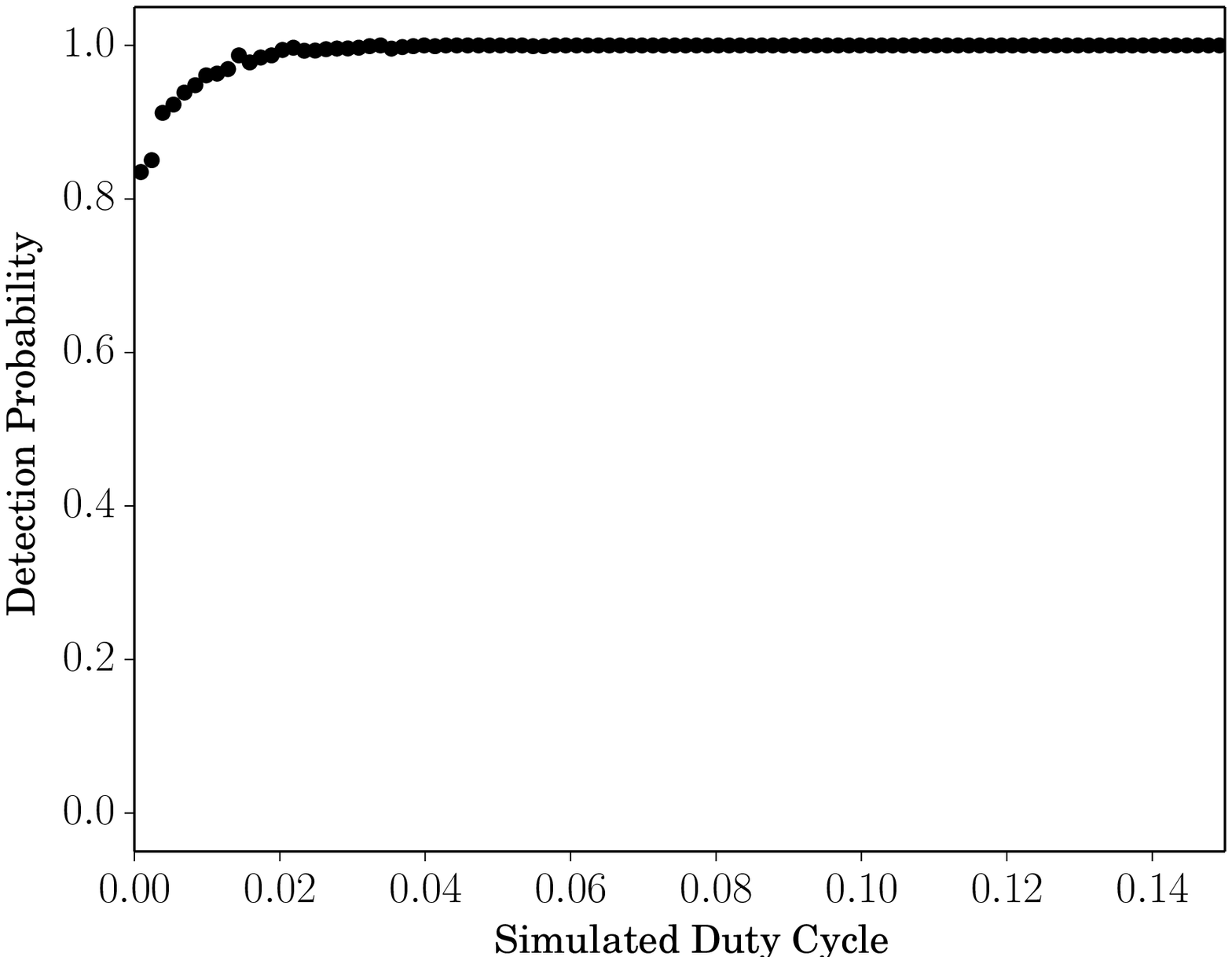}
\includegraphics[scale=0.47,viewport = 0 0 530 410, clip]{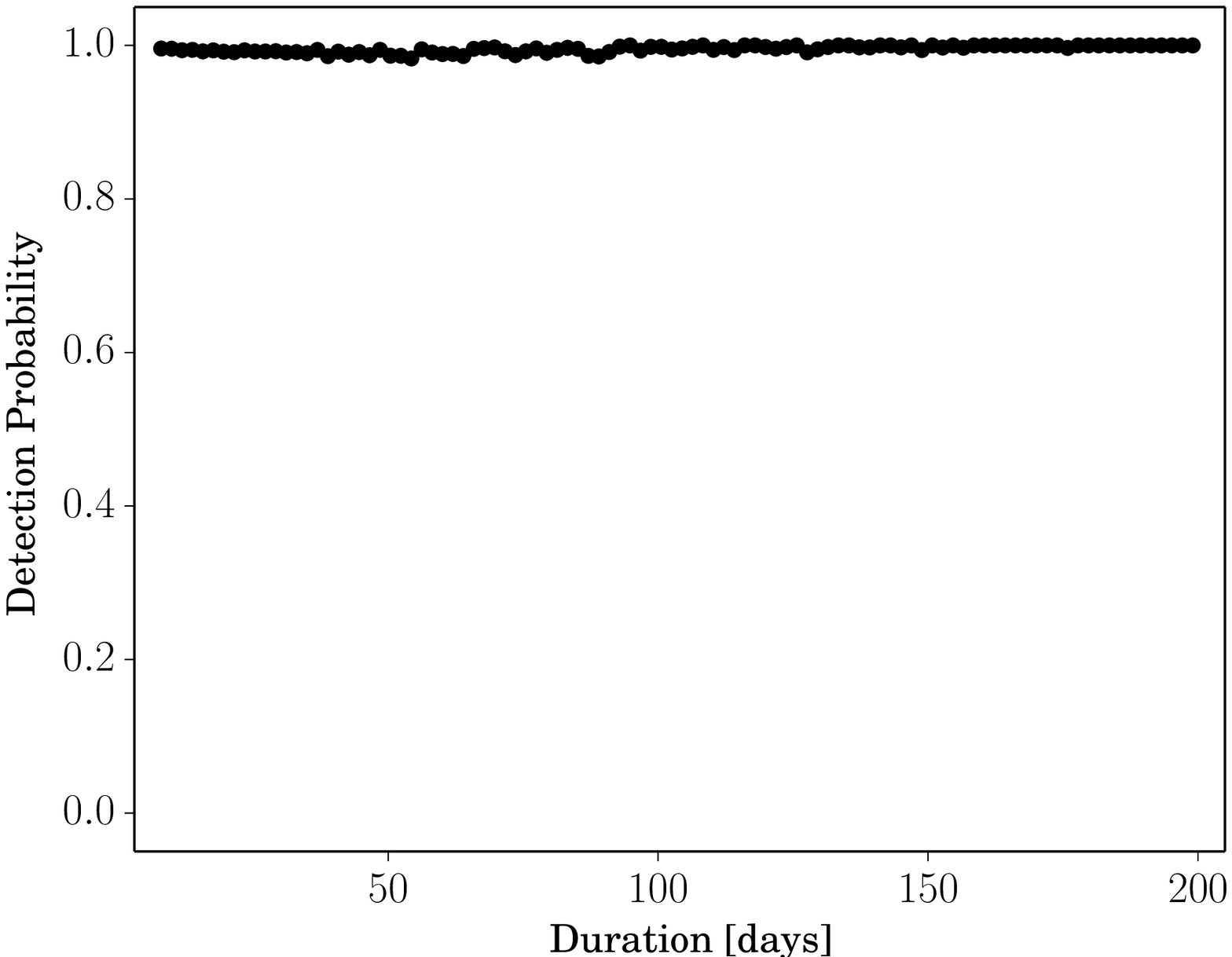}
\includegraphics[scale=0.46,viewport = 0 0 555 410, clip]{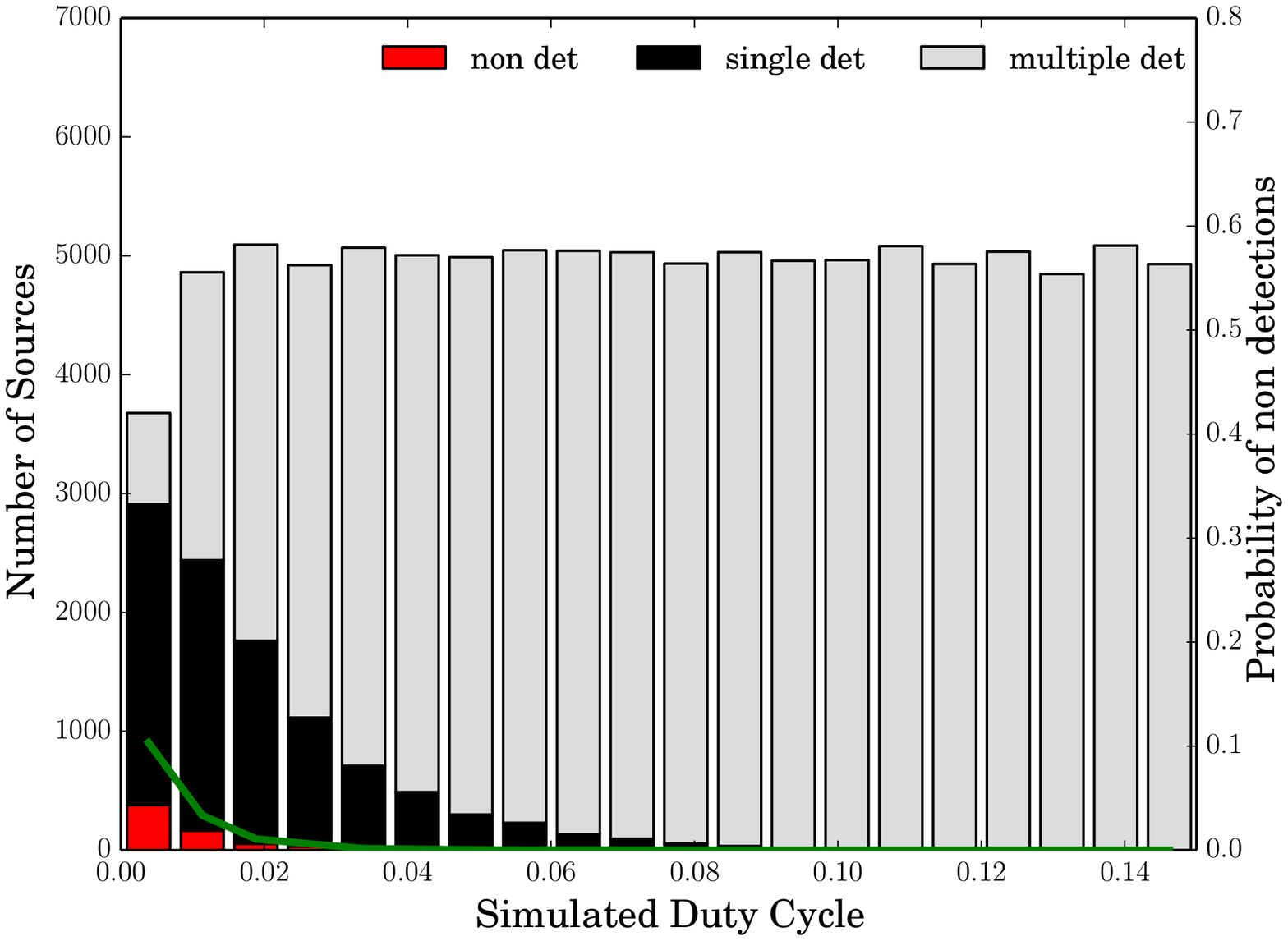}
\includegraphics[scale=0.47,viewport = 0 0 520 410, clip]{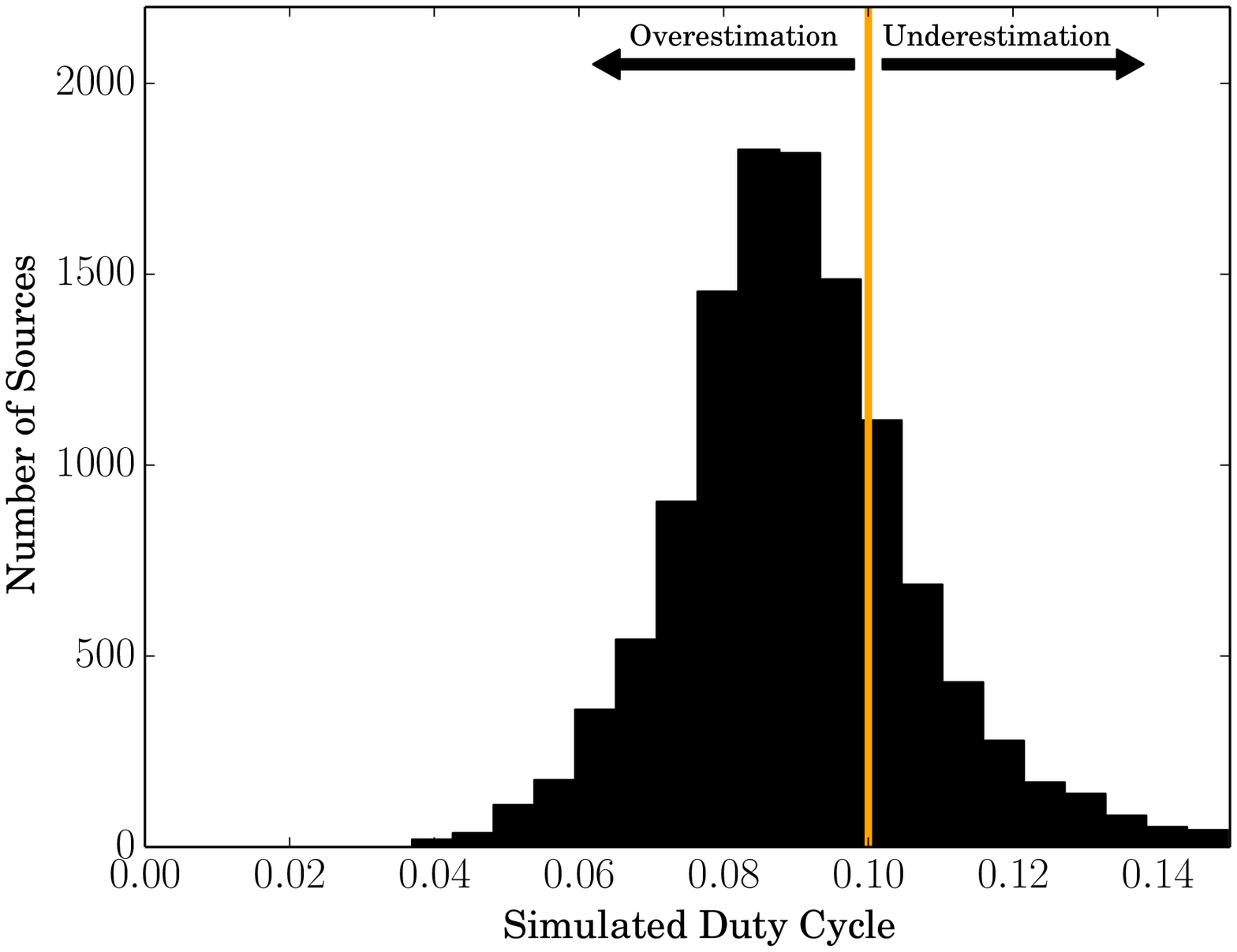}
\includegraphics[scale=0.435,viewport = 0 0 570 410, clip]{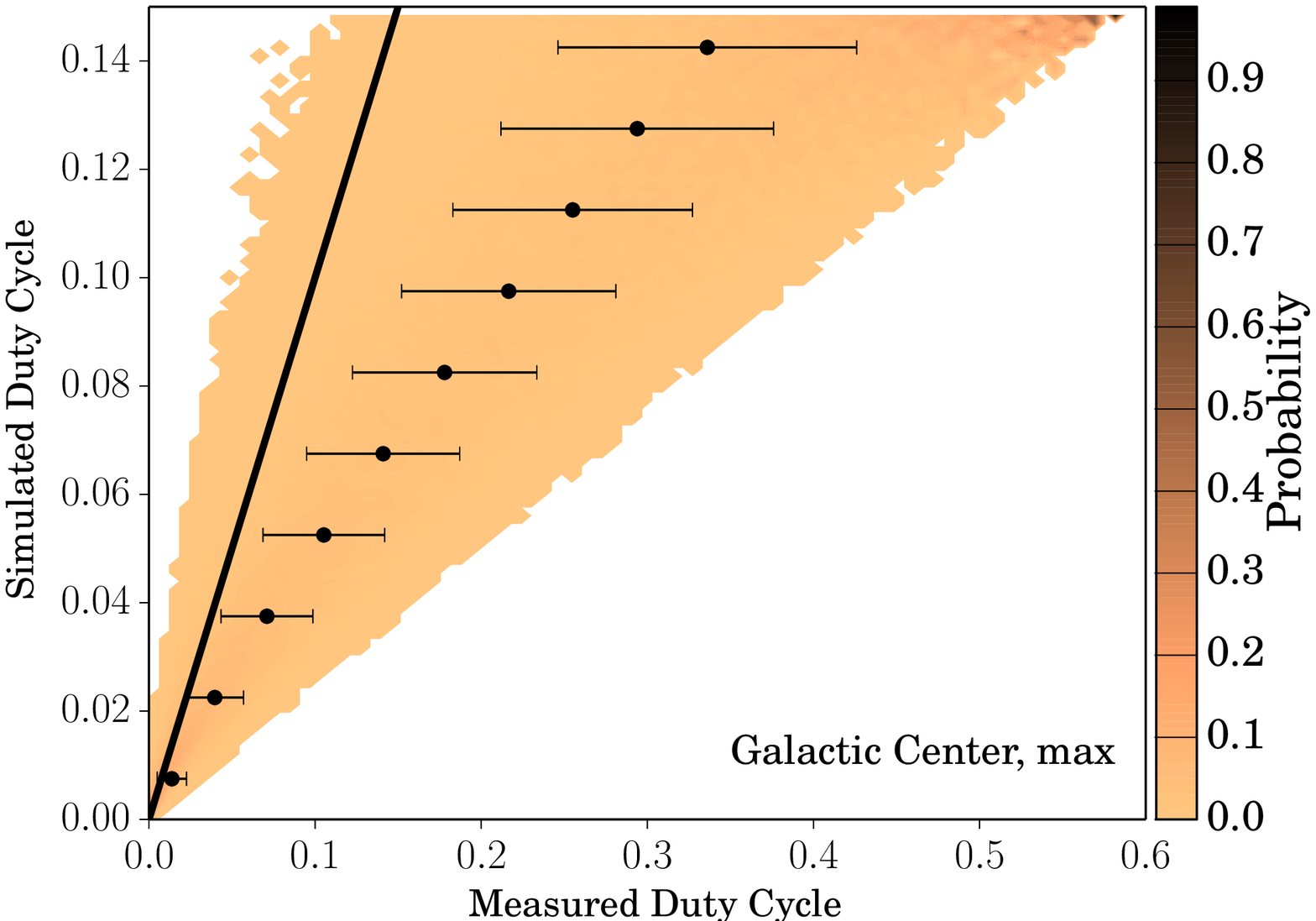}
\includegraphics[scale=0.435,viewport = 0 0 570 410, clip]{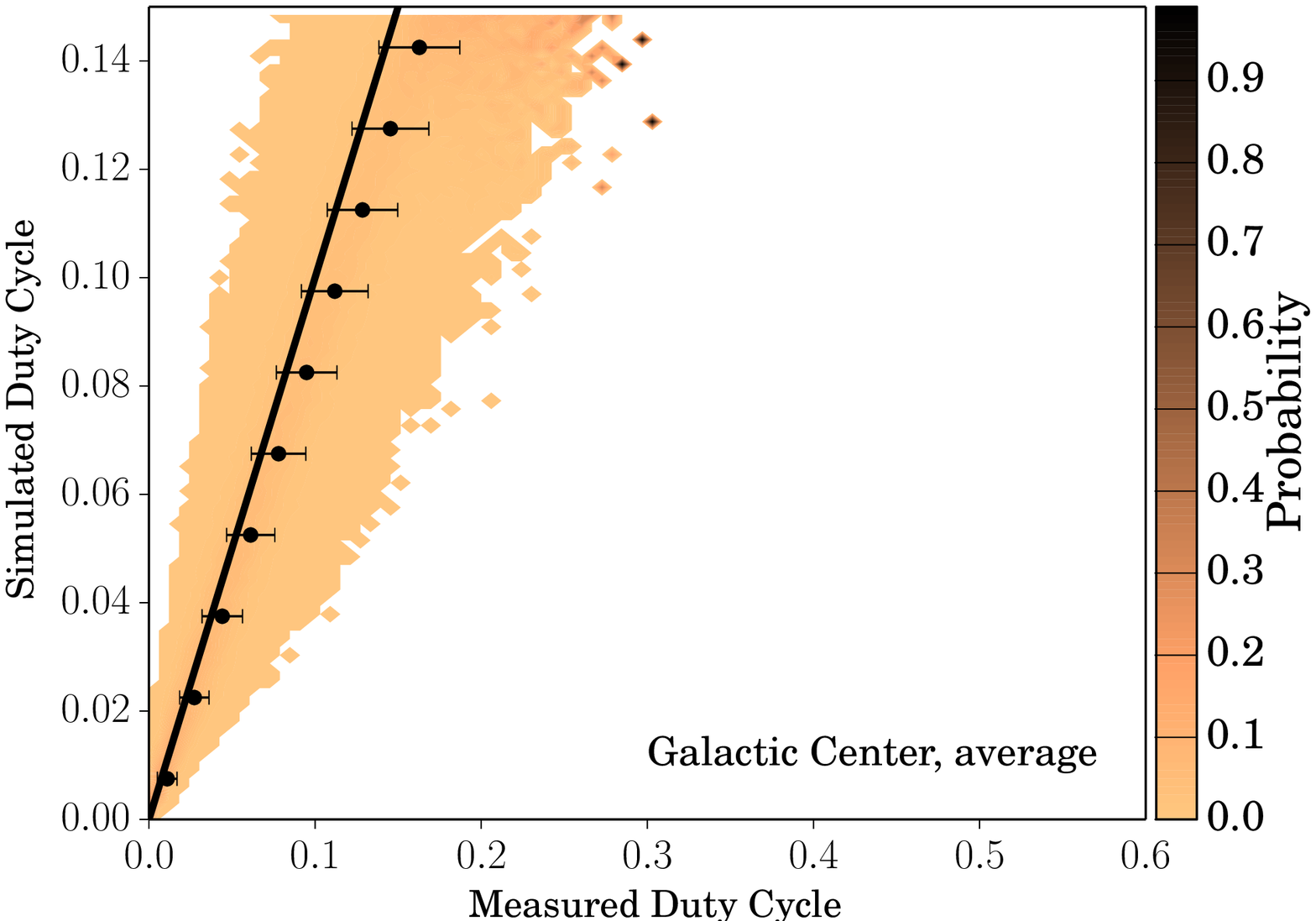}
\caption{
Results of our simulations using the observing campaign of the Galactic Centre as input observing campaign.
The six panels are the same as in Figure~\ref{fig:ngc6388}.
}
\label{fig:GC}
\end{figure*}

\subsection{The Galactic centre}
\label{sec:GC}
In Figure~\ref{fig:GC} we show the results of our simulations using the \emph{Swift}/XRT observing campaign of the Galactic centre.
This survey combines the positive things of the previous two: it lasted for more than a decade, enabling many sources to have
multiple detected outbursts, and had very frequent observations, without long gaps (apart from the Solar impediments),
avoiding most outbursts to fall in these periods.

As can be seen from the top left panel of this figure, most sources are detected. Only sources with simulated DC smaller than
$\sim$\,0.02 have about 10 to 20 percent probability of not being detected.
This implies that all the transients near the Galactic centre with DC higher than 0.02, which underwent an outburst during the observing
campaign, must have been detected (within our assumptions on the sources properties such as outburst duration). 
Moreover, if a new source exhibits its first outburst after the current observing campaign, it can have a maximum DC of
D$_{\textrm{max}}$\,=\,max(T$_{\textrm{outburst}}$) / T$_{\textrm{survey}}$ = 0.05, where T$_{\textrm{survey}}$ is the duration of the survey
($\sim$\,10 years). This implies that all the transient sources which exhibit outbursts with durations between 7 and 200 days and have
a DC larger than 0.05 have been detected already. 

We note that sources that exhibit very long outbursts (e.g., $>$\,1 year; the so-called quasi-persistent
sources; see footnote 2) and have DC $<$\,0.1 would have quiescent periods $>$\,10 years and therefore they could have still
remained undetected by this survey.
Another type of transient that might be missed with this survey campaign would be the one having periodic outbursts with
recurrence time almost exactly multiple of a year, with outbursts all coinciding with the gap in the observations due to Solar
constraints, and duration shorter than that gap. 

In the top right panel of Figure~\ref{fig:GC} we can see that all outbursts have close to 100 percent probability of being detected regardless
of their duration. This means that DC is the only variable playing a role in the detectability.

As mentioned, most sources are detected multiple times with this strategy, and very few outbursts are actually missed.
This reflects in the middle left and the two bottom panels of Figure~\ref{fig:GC}.
In the middle left panel we can also see that the probability of a source not being detected at all is never higher than 10\%.
In the two bottom panels, if we estimate {\textrm DC}$_{\textrm{obs}}$ as the maximum,
on the left, we largely overestimate {\textrm DC}$_{\textrm{obs}}$ as our measurements are biased by the fluctuations in DC that
we simulates (as described in Section~\ref{sec:methods}). On the other hand, if we use the average value, we only slightly
overestimate {\textrm DC}$_{\textrm{obs}}$, with quite narrow error bars.
This is confirmed also looking at the middle right panel, where we clearly see that for most sources
with {\textrm DC}$_{\textrm{obs}}$\,=\,0.10 we are overestimating their DC by about 10 percent.

\begin{figure*}
\centering
\includegraphics[scale=0.40,viewport = 0 0 570 410, clip]{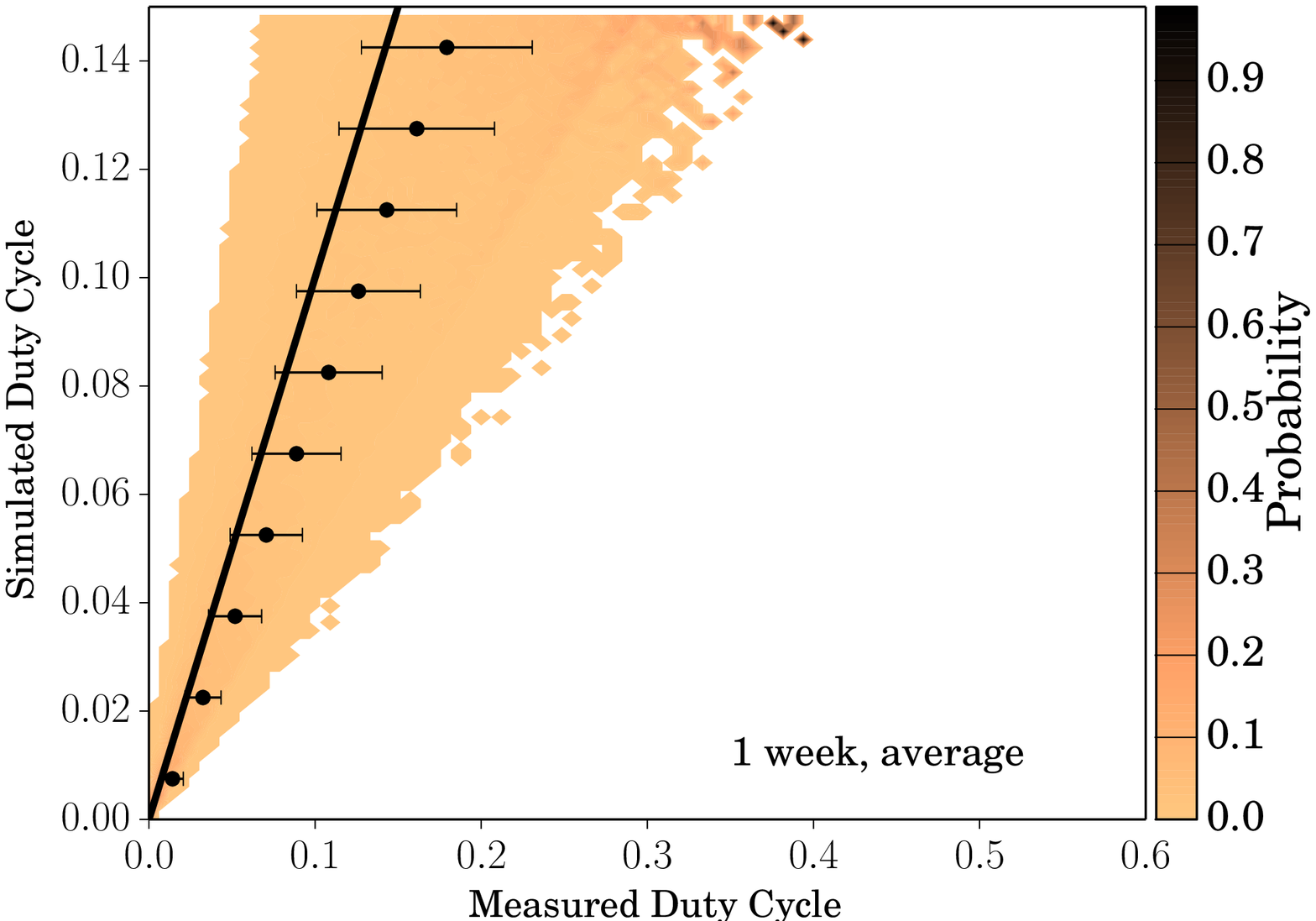}
\includegraphics[scale=0.40,viewport = 0 0 555 410, clip]{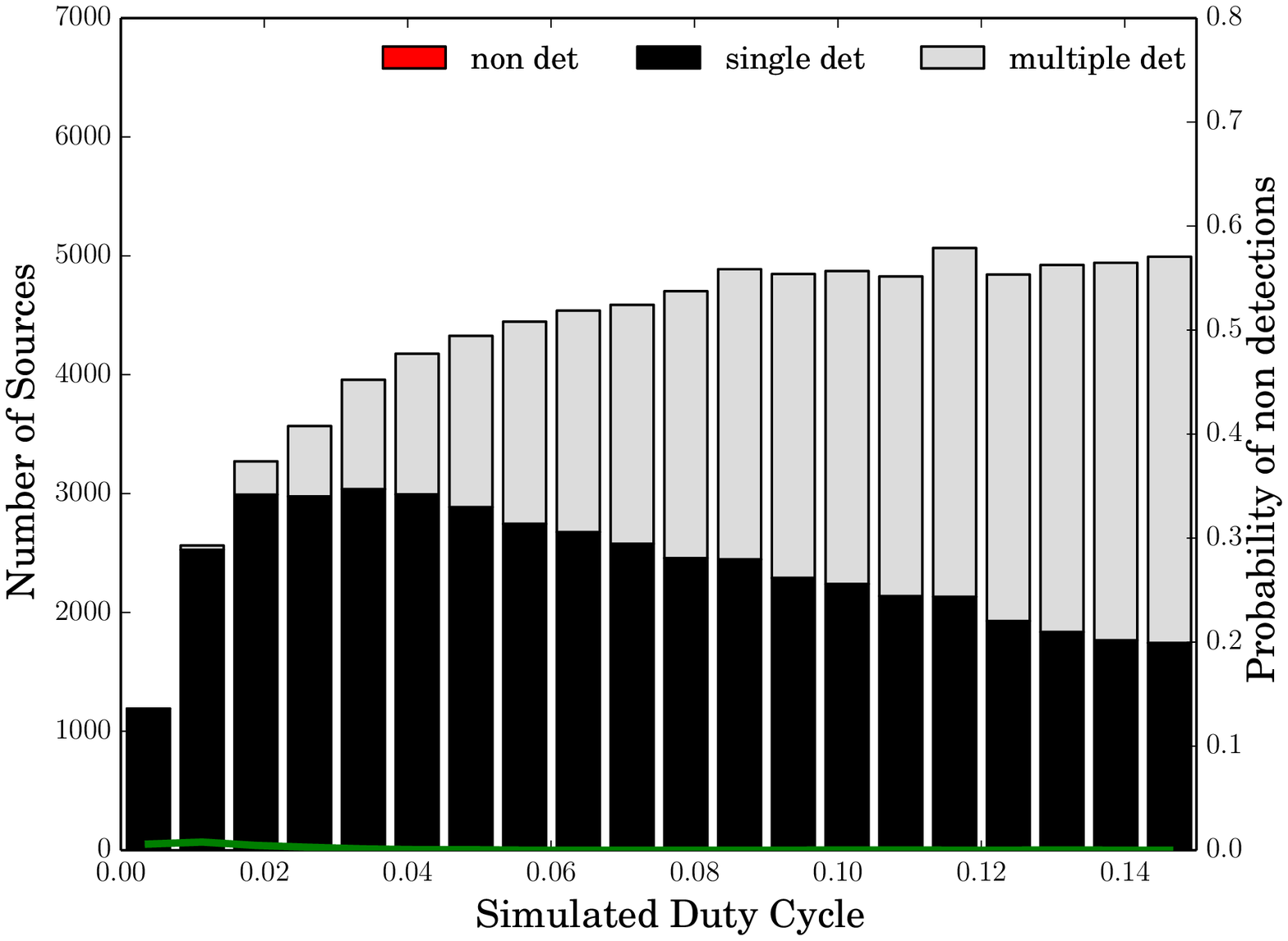}
\includegraphics[scale=0.40,viewport = 0 0 570 410, clip]{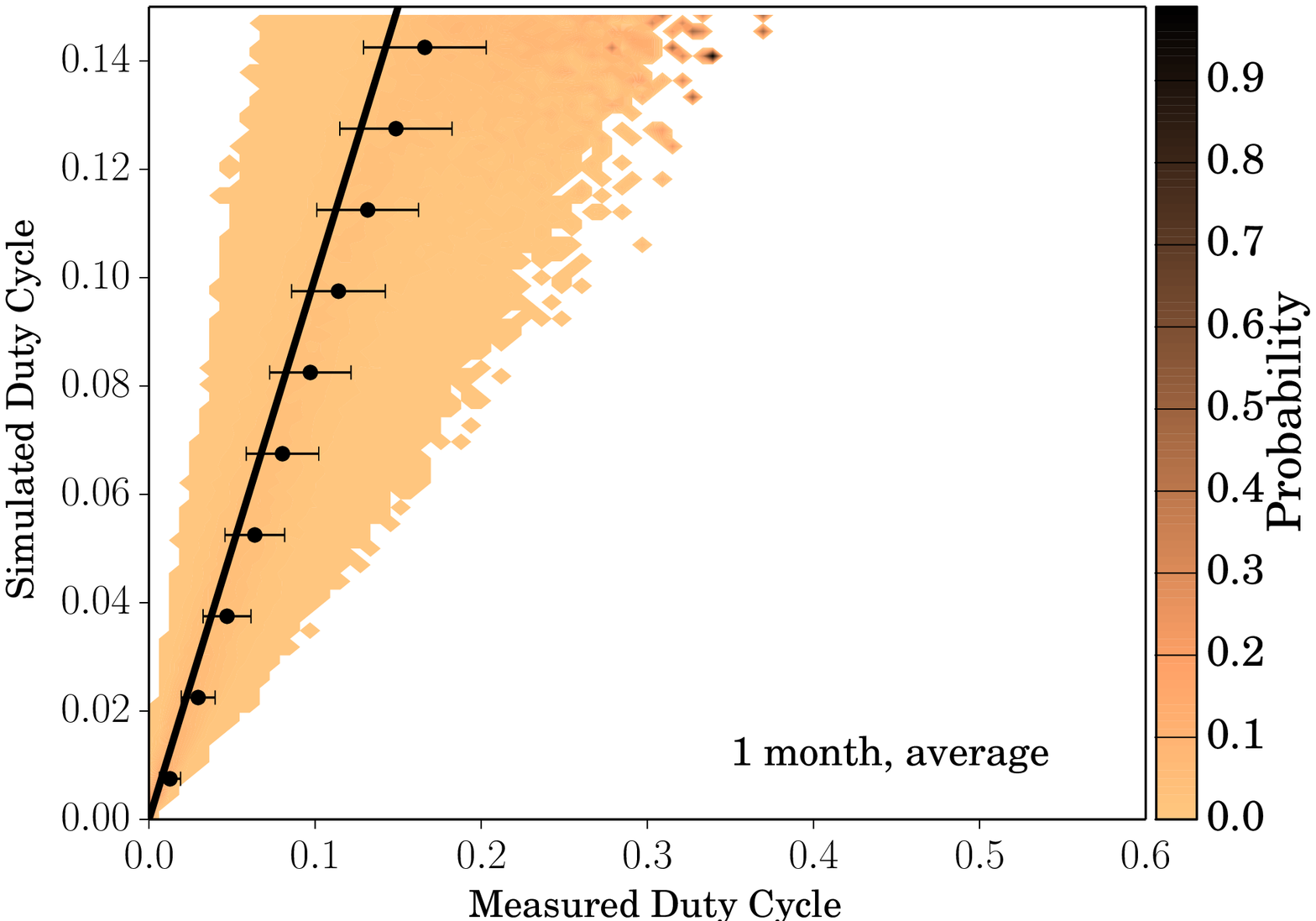}
\includegraphics[scale=0.40,viewport = 0 0 555 410, clip]{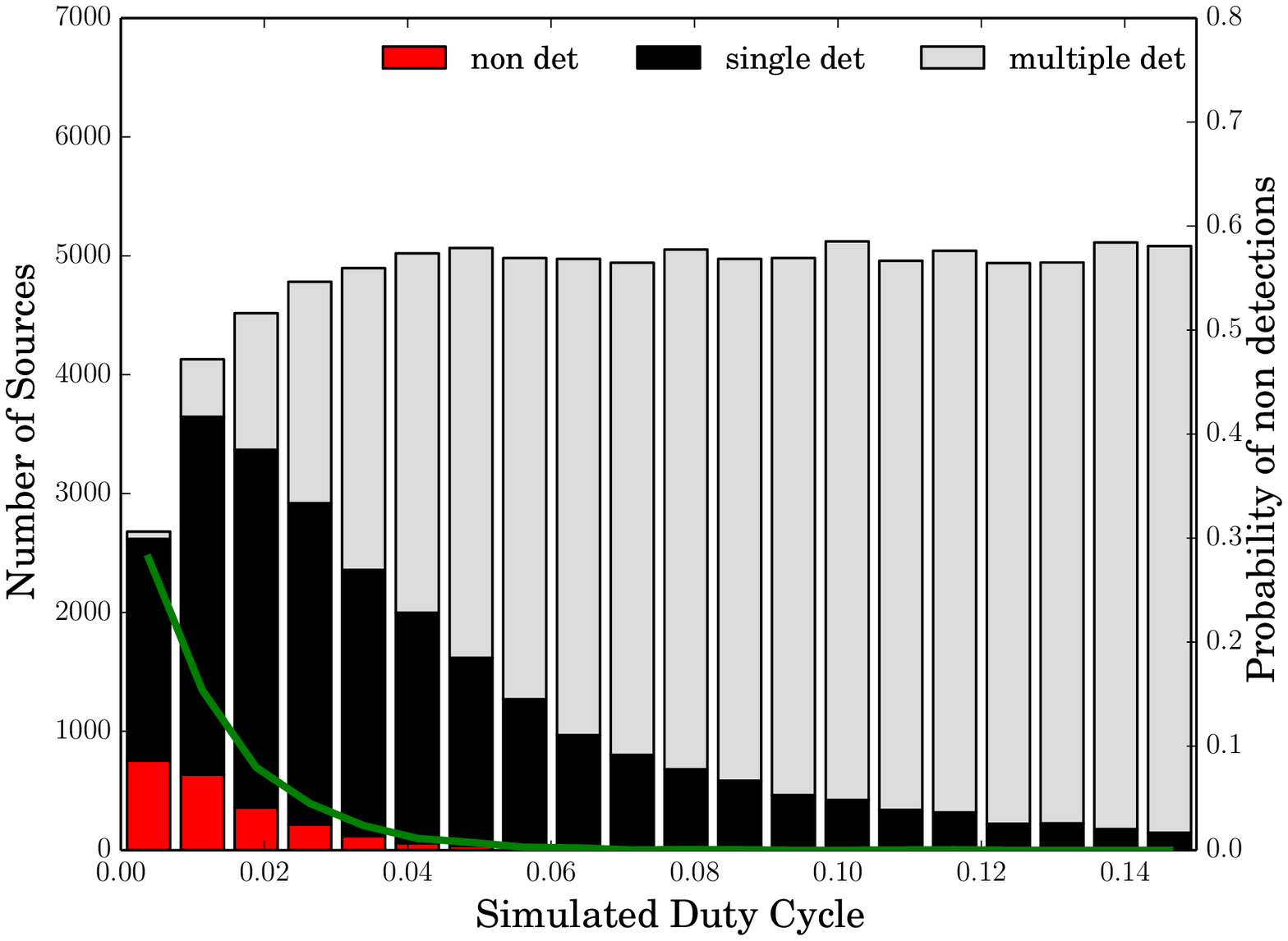}
\includegraphics[scale=0.40,viewport = 0 0 570 410, clip]{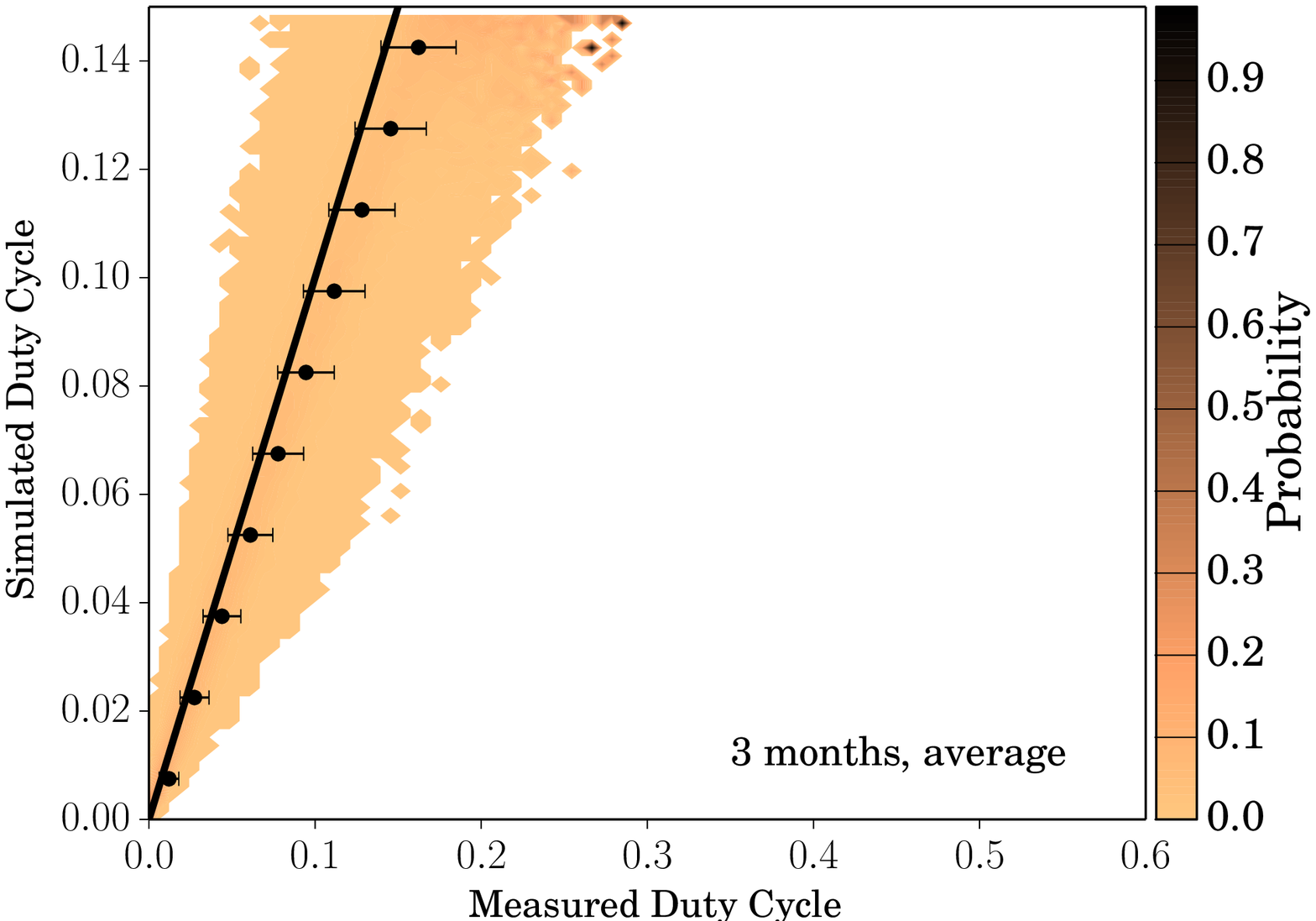}
\includegraphics[scale=0.40,viewport = 0 0 555 410, clip]{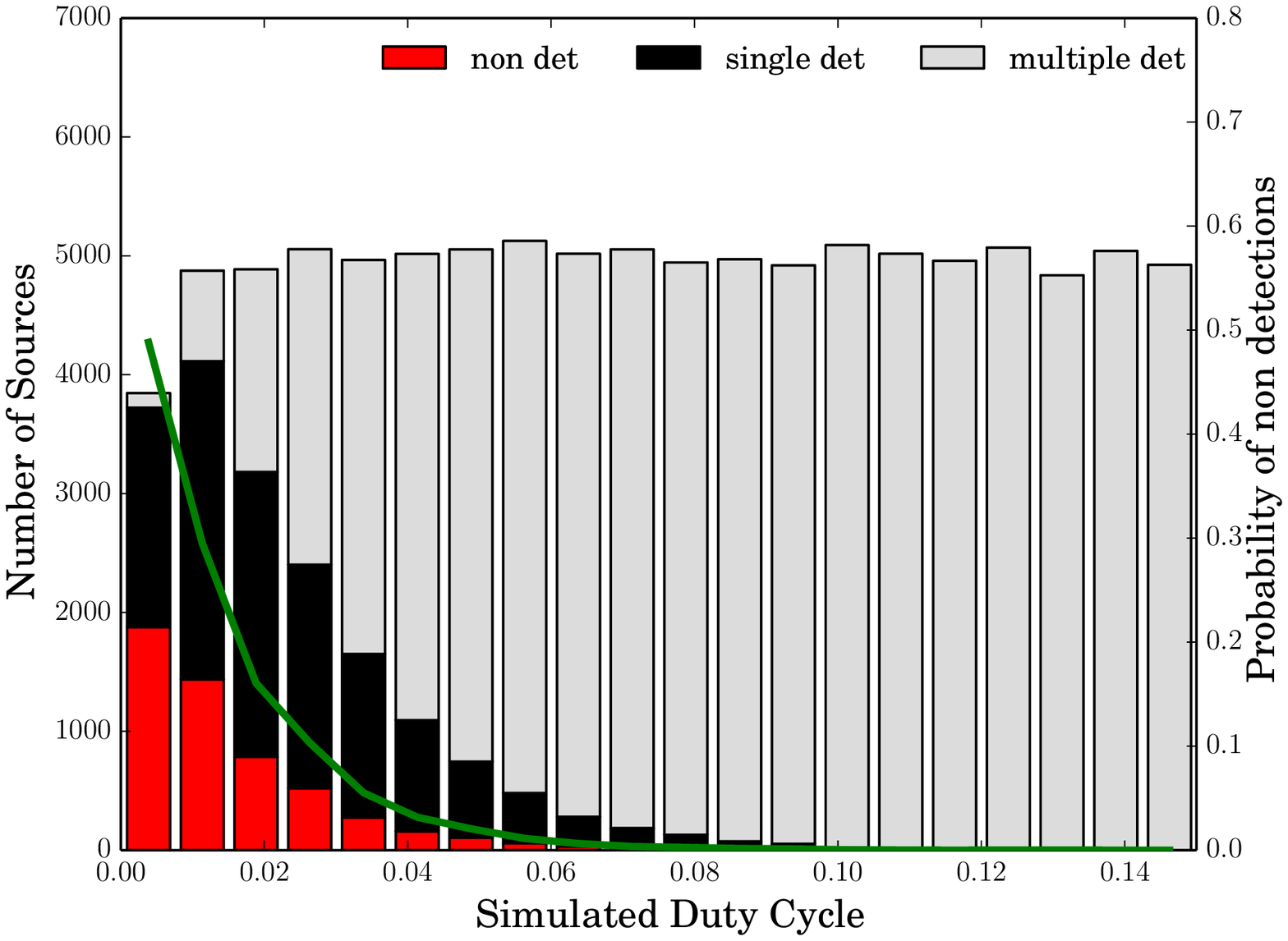}
\includegraphics[scale=0.40,viewport = 0 0 570 410, clip]{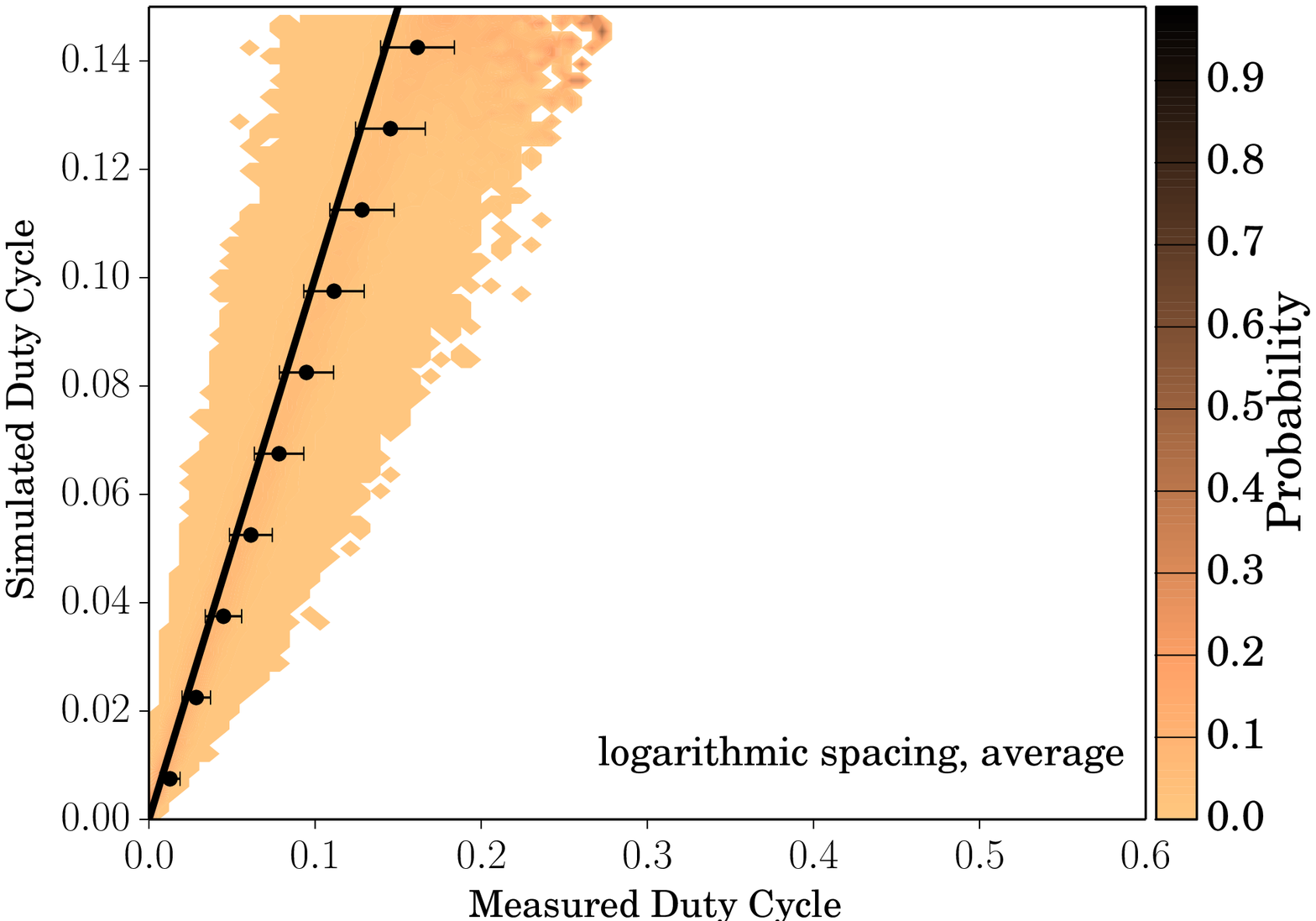}
\includegraphics[scale=0.40,viewport = 0 0 555 410, clip]{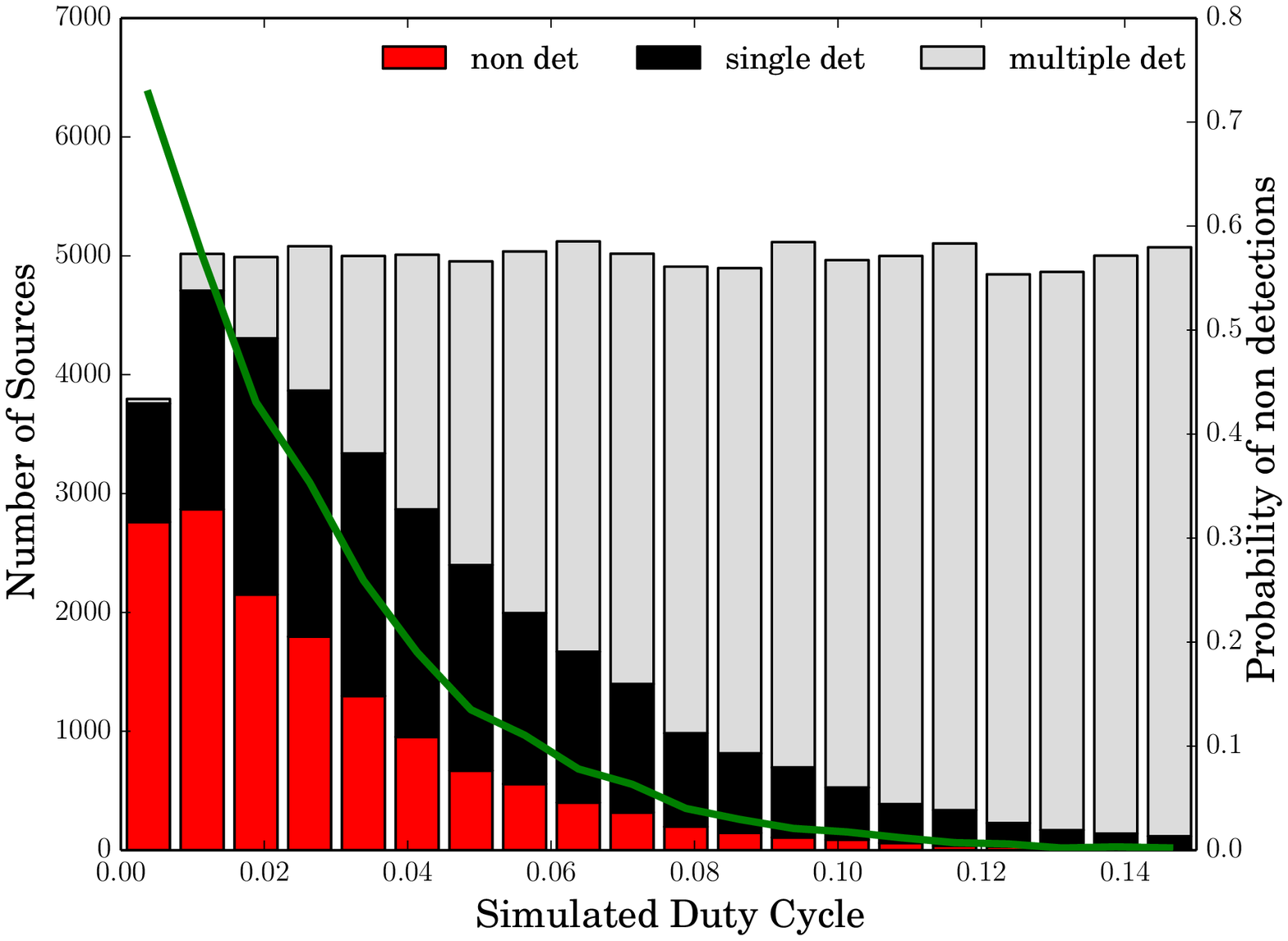}
\caption{
Plot of the bottom right and middle left panels as in Figure~\ref{fig:ngc6388} for the four artificial survey strategies described
in Section~\ref{sec:artificial}. All of the strategies are composed of 58 observations of 1\,ksec.
From top to bottom, the observations are one week, one month, three months apart, and logarithmically spaced.
}
\label{fig:artificial}
\end{figure*}

\subsection{Different Survey Strategies}
\label{sec:artificial}
We have performed simulations for several artificial observing campaigns to determine what kind of strategies would optimize
the detection of transients, and would result in the most accurate estimation of their DCs. For the total observing time of those artificial
strategies we have chosen the one of the Terzan~5 campaign (total observing time of $\sim$\,58 ksec)
as a representative sample for what can be obtained with \emph{Swift}/XRT. We divided this in 58 observations of equal duration of
1\,ksec.

\begin{figure*}
\centering
\includegraphics[scale=0.44,viewport = 0 0 530 410, clip]{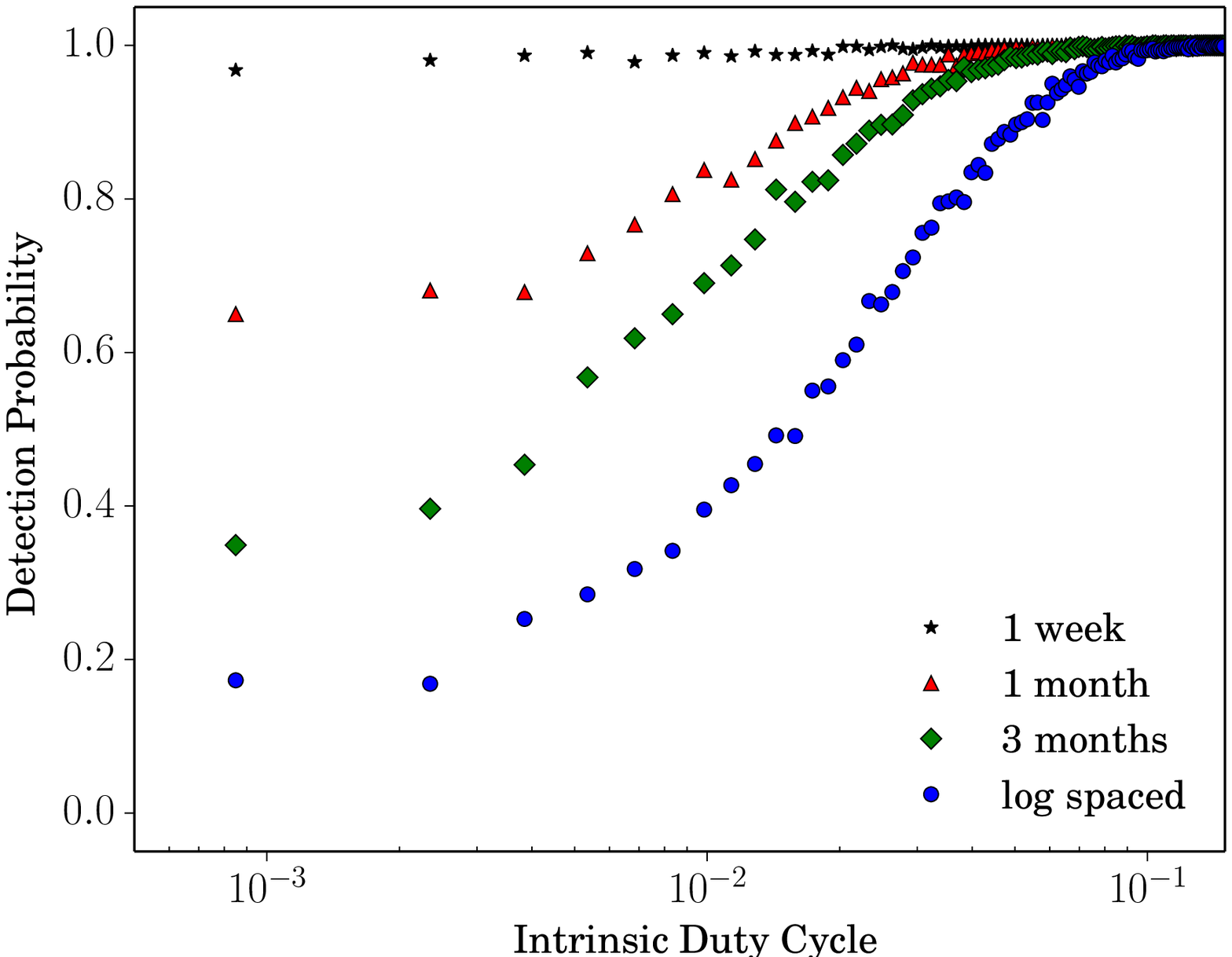}
\includegraphics[scale=0.44,viewport = 0 0 530 410, clip]{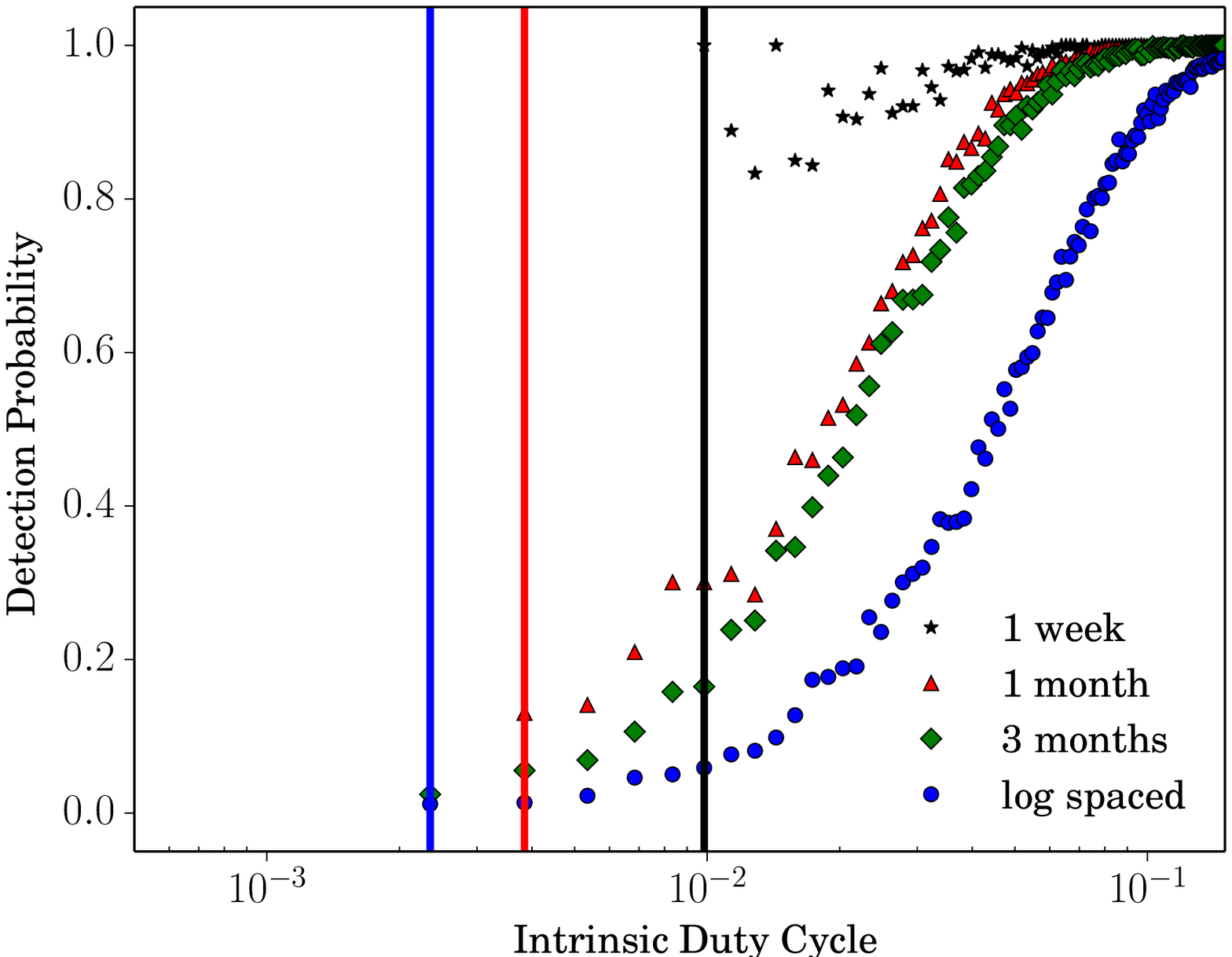}
\includegraphics[scale=0.44,viewport = 0 0 530 410, clip]{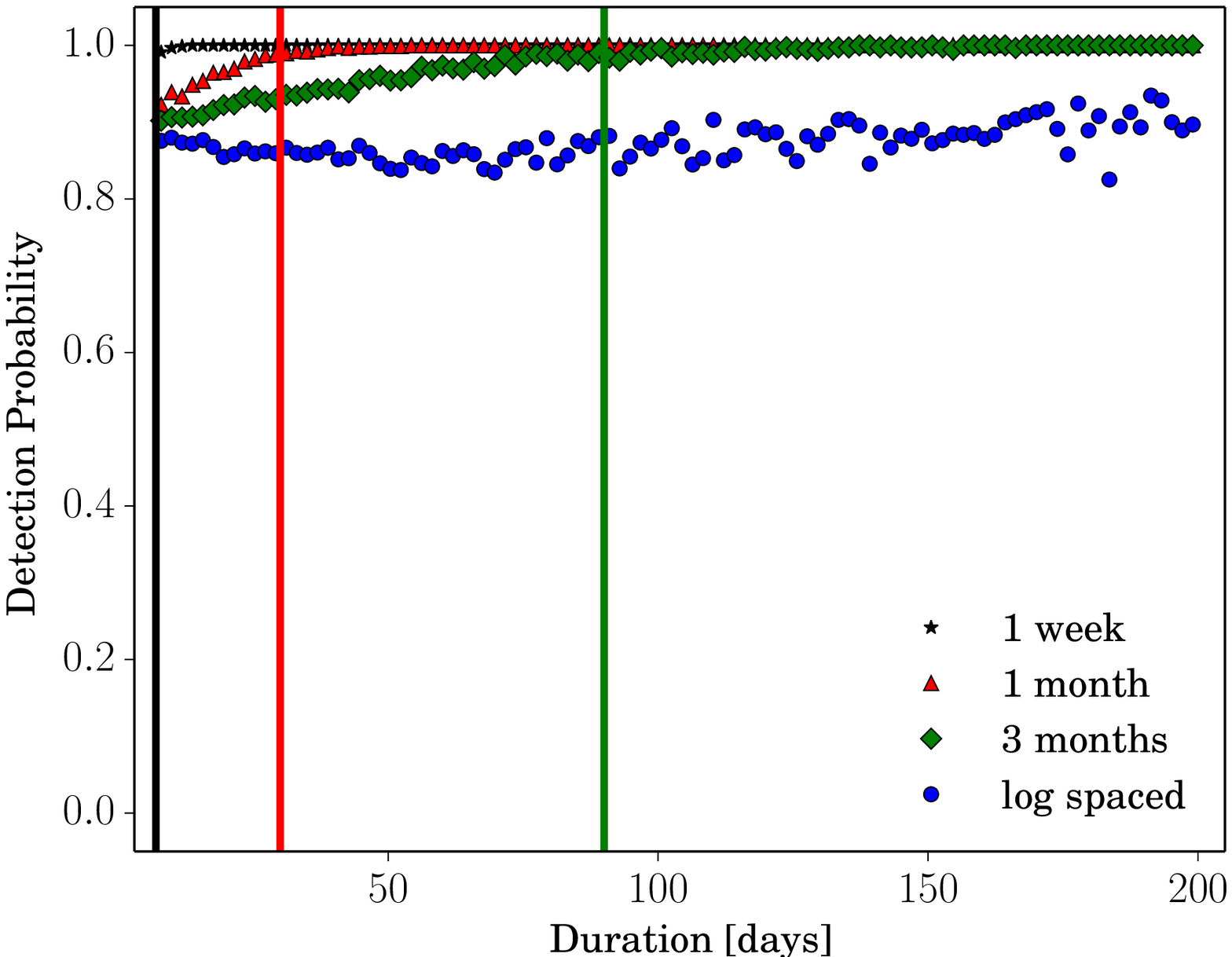}
\caption{
Plot of the probability of detecting simulated sources as a function of their simulated DC, for all sources in the top left panel, and
only for sources exhibiting multiple outbursts during the observing campaign in the top right panel. The vertical lines in the right panel
represent the minimum DC different artificial surveys can measure.
The bottom panel shows the probability of detecting outbursts as a function of their duration. The vertical lines in the bottom panel
represent the longest gap between two consecutive observations in different artificial surveys.
}
\label{fig:combined}
\end{figure*}

We then spaced these observations in different ways: one week apart (similar to the strategy used for NGC~6388), one month
apart, three months apart and, logarithmically spaced.
For the last strategy, we divided the observations in blocks of 7 observations. The observations within the same block are
uniformly spaced, but separated by different amounts in different blocks. The separations we used are 1 day, 4 days, 7 days,
14 days, 1 month, 3 months, 6 months and 1 year. The separation between the last observation of a block and the first of the
following is set equal to the separation between observations of the latter. While the total exposure time is constant, the length
of the observing campaign is different for each of the strategies, ranging between 398 days ($\sim$\,1\,year) for the most compact
one and 5292 days ($\sim$\,14.5\,years) for the logarithmically spaced strategy. The results of our simulations are shown in
Figures~\ref{fig:artificial} and \ref{fig:combined}.

In the left panels of Figure~\ref{fig:artificial} we show the same plot as in the bottom right panel of Figure~\ref{fig:ngc6388}, while in
the right ones we show the same plot as in the middle left panel of Figure~\ref{fig:ngc6388}.
Observing the left panels in Figure~\ref{fig:artificial}, we note that in all cases {\textrm DC}$_{\textrm{sim}}$ is slightly overestimated
and that the error bar we find shrink for longer and longer campaigns.
In the right panels of Figure~\ref{fig:artificial} we can instead observe that going from spacing of 1 week to 1 month,
and then to 3 months we are able to detect more and more sources and more and more sources are detected multiple times
(i.e., we were able to measure their {\textrm DC}$_{\textrm{obs}}$).
The logarithmically spaced strategy is constituted by an initial cluster of very close spaced (in time) observations that become
less and less frequent, and in the second half of the survey observations are a year apart. This causes many outbursts to be
undetected and therefore many sources with a small {\textrm DC}$_{\textrm{sim}}$ are not detected because their few (or only) outbursts
end up during the very long gaps in this campaign. These effects are clear in the bottom right panel of Figure~\ref{fig:artificial}.
We also note a dramatic change in the probability of completely missing a source as a function of its duty cycle. This probability
is extremely small with observations spaced 1 week from one another ($<$1\%), it reaches about 30\% when they are 1
month apart, 50\% if they are 3 months apart, and 75\% in case they are logarithmically spaced.

We compared the results of these surveys in Figure~\ref{fig:combined}.
In the left panel we show the probability of detecting a source as a function of its {\textrm DC}$_{\textrm{sim}}$
for different artificial strategies.
We note that the most compact strategy (marked with black stars) can detect basically all sources undergoing an outburst
during the observing campaign. The probability of detecting sources decreases steadily
as the gap between consecutive observations increases from 1 week, to 1 month, to 3 months.
The survey strategy with observations logarithmically spaced is the one that has the lowest probability of detecting sources.
This is due to the fact that it has very large gaps between consecutive observations in the latter part of the campaign.
This is the reason why dense campaigns are at an advantage.

In the right panel of Figure~\ref{fig:combined} we show the same as on the left panel, but restricting the source sample to
only those that had two or more detected outbursts during the observing campaign.
Here it is clear that each survey can observe multiple outbursts from sources only down to a limit DC that is related to the
total length of such survey. The smallest DC a survey can observe is given by the ratio between the shortest outburst duration
and the total length of the survey. This is the reason why long campaigns are at an advantage.

In the bottom panel of Figure~\ref{fig:combined} we show the probability of detecting outbursts as a function of their duration.
We can see that each survey has a probability of 1 of detecting outbursts longer than the longest gap in them.
Outbursts shorter than that are only occasionally detected, and the probability of detecting such outbursts is directly proportional
to their duration, as shown in \citet{Carbone2017}.

\subsection{Single Outbursts Detections}
\label{sec:single_det}
\begin{figure}
\centering
\includegraphics[scale=0.44,viewport = 0 0 570 410, clip]{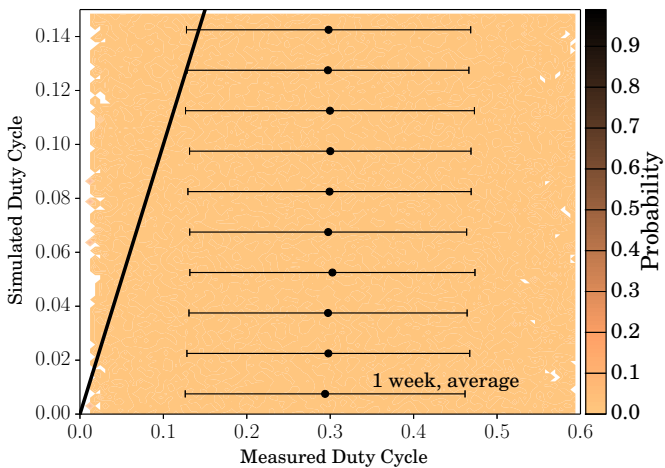}
\caption{
Same plot as in the left panels in Figure~\ref{fig:artificial} for the artificial strategy with observations one week apart, but
plotting only sources with only one detected outburst. It is clear that in this case we cannot gather any information
on {\textrm DC}$_{\textrm{sim}}$ given we estimated {\textrm DC}$_{\textrm{obs}}$ from a single outburst detection.
}
\label{fig:singles}
\end{figure}

Finally, we have tested whether we could constrain in any way the DC of sources for which only one outburst was
detected in our simulations using the artificial survey strategies as discussed in section \ref{sec:artificial}. 
Our approach to this problem is the following: for all sources with a single detected outburst, we have calculated the time at
which an outburst with the same duration could have started and not have been detected. This could be either before our campaign
started, after the campaign ended, or in a gap long enough during the campaign itself. In all three cases we calculated what the
corresponding DC would have been, and we selected the largest obtained value.
We have estimated the duration of a source as the time difference between the first and last detection.
An example of the results from this analysis is shown in Figure~\ref{fig:singles} for the strategy in which the observations are
spaced by one week. It is clear that provided we estimate a certain value for {\textrm DC}$_{\textrm{obs}}$ in case of a single detection,
this does not allow us to reconstruct the simulated value of the DC.

\section{Discussion and Conclusions}
\label{sec:conclusion}
Using an expanded version of the transient simulation code of \citet{Carbone2017},
we have simulated the X-ray light curves of outburst from transient LMXBs to investigate the bias that an observing campaign
can introduce in the calculation of DC of these sources, and in particular we focussed on the VFXTs.

In our simulations we used as input survey strategy, the \emph{Swift}/XRT observing campaigns of the globular clusters NGC~6388 and 
Terzan~5, and the very extensive campaign on the Galactic centre. Those campaigns were chosen because they give us a good
variety in density of the observation sampling, and the total duration of the campaigns. Therefore, our results should be directly
applicable to those, and to similar observational strategies. From our simulations of the survey of the Galactic centre, we
determined that all the transient LMXBs in that region, with DC larger than 0.02, undergoing at least an outburst 
during the observing campaign have been detected.
Moreover, if a new source will exhibit its first outburst since the beginning of the monitoring
campaign, it will have a DC smaller than 0.05, if the duration of the previous outburst was smaller than 200 days.
This implies that all the transient sources with outbursts shorter than 200 days, and DC higher than 0.05 have been
already detected. 
However, quasi-persistent sources, which have very long outbursts ($>$\,1 year), could still have remained
undetected despite they might have higher DCs, because of their very long quiescence period that could extend longer than the
campaign has been active.
Another type of transient that might be missed with this campaign would be the one having periodic outbursts with
recurrence time almost exactly multiple of a year, with outbursts all coinciding with the gap in the observations due to Solar
constraints, and duration shorter than that gap.

From our simulations it is clear that fluctuations in outburst duration and recurrence times affect our estimation of the DC more than
non detected outbursts. This biases our measures to overestimate the simulated DC of sources.
The next step in such simulations is to model fluctuations in both the outburst duration and the recurrence time with Gaussian
distributions.
Since real transients have also a variation in their outburst duration and the duration of their quiescence period 
\citep[e.g., see][]{Yan2015}, determining the DC of those transients \citep[see][for the DC of VFXTs and \citealt{Yan2015} for
the DC of the brighter transients]{Degenaar2010} will very likely also suffer this bias.
We note that despite we performed our simulations with a focus on VFXTs, very likely this conclusion
(i.e., DC calculations being affected more by fluctuations in outburst duration and recurrence times rather than
undetected outbursts) applies for brighter transients as well.

From our analysis of the probability of detecting individual sources, we have determined that
compact surveys are necessary to detect outbursts with short durations because we showed that a survey is detecting all sources
with duration longer than the maximum separation between consecutive observations,
while the detection probability decreases for shorter and shorter outbursts.
On the other hand, long surveys are necessary to detect sources with low DC because the smallest DC a survey can observe is given
by the ratio between the shortest outburst duration and the total length of the survey.
If one has a limited amount of observing time these two effects are competing, and a compromise is required which is set by the goals
of the proposed survey.

In order to investigate what the best observing campaign would be to maximize the probability to detect transients as well as to have
the most accurate estimates of the observed DC, we have also performed simulations with several different artificial survey
strategies (see Section~\ref{sec:artificial} for details). As expected, the best campaign would be a regular monitoring that extends
for a very long period, without any long gap between observations. The closest real example of such a dataset is the monitoring
of the Galactic Centre.

We have simulated artificial surveys with regular separations between consecutive observations of 1 week, 1 month and
3 months. We have also simulated a survey composed of blocks with observations logarithmically spaced.
As expected, we found that the survey with observations 1 week apart is the one that give the highest probability of
detecting individual sources (i.e., detecting at least one outburst from a source if it was active during the observing campaign,
see Figure~\ref{fig:combined}).
We determined that the survey with logarithmically spaced observations is the one that has the lowest probability of detecting
transients, despite it can probe lower values of the simulated DC. Such survey resembles strategies in which a target field was
observed with a dense sampling for a certain period and then it gets a few sparse observations later on.
We have shown how such approach might not lead to any further constraints on the DCs, nor increase significantly the likelihood
of detecting new transients.
A better strategy would be to initiate a new dense monitoring of the same region rather than have few individual points per year.

We have shown that the minimum DC that can be determined using a specific survey is a function of its total duration,
as longer surveys can probe lower DCs, and as most of the DCs for real transients are below 0.1, a survey
of duration of at least months is required to probe that regime. We have also proved that very dense surveys, with observations
every a few days, will not miss (almost) any outburst and will therefore not be affected by the issue of underestimating the DC,
assuming that the observing campaign lasts at least one full cycle (outburst plus quiescence).

An expansion to our simulation code would be to also include the calculation of $\langle\dot{M}_{\textrm A}\rangle$. As explained in the
introduction, accurate $\langle\dot{M}_{\textrm A}\rangle$ is very important for binary evolution and population models, as well as to
study certain types of neutron star physics. Finally, another possible development of our
simulations would be the inclusion of different distributions of the variations in DC and outburst duration.

\section*{Acknowledgments}
D.C. acknowledges support from NSF CAREER award \#1455090.
We thank Nathalie Degenaar and Ralph Wijers for comments on an earlier version of this paper.
This research has made use of NASA's Astrophysics Data System Bibliographic Service.

\appendix
\section{List of the Observations}

In Table~\ref{tab:observations} we report the dates of all the observations we used in our simulations regarding the campaigns
on Terzan~5 and NGC~6388.

\begin{table}
\begin{center}
\begin{tabular}{|ccc|c|}
\hline
\multicolumn{3}{c|}{Terzan~5}	& NGC~6388 \\
\hline
2010-10-28 & 2015-03-21 & 2016-04-21 & 2012-06-02 \\
2010-10-31 & 2015-03-23 & 2016-04-23 & 2012-06-09 \\
2011-10-26 & 2015-03-25 & 2016-04-29 & 2012-06-16 \\
2012-02-09 & 2015-03-26 & 2016-04-30 & 2012-06-23 \\
2012-06-11 & 2015-03-27 & 2016-05-01 & 2012-06-30 \\
2012-06-16 & 2015-04-02 & 2016-05-15 & 2012-07-08 \\
2012-06-21 & 2015-04-06 & 2016-05-29 & 2012-07-14 \\
2012-06-26 & 2015-04-12 & 2016-05-30 & 2012-07-18 \\
2012-06-30 & 2015-04-13 & 2016-06-12 & 2012-07-21 \\
2012-07-06 & 2015-04-14 & 2016-06-26 & 2012-07-28 \\
2012-07-07 & 2015-04-20 & 2016-06-29 & 2012-08-08 \\
2012-07-08 & 2015-04-22 & 2016-07-10 & 2012-08-11 \\
2012-07-10 & 2015-04-24 & 2016-07-13 & 2012-08-18 \\
2012-07-12 & 2015-04-26 & 2016-07-24 & 2012-08-25 \\
2012-07-13 & 2015-04-28 & 2016-08-07 & 2012-09-01 \\
2012-07-16 & 2015-05-02 & 2016-08-21 & 2012-09-08 \\
2012-07-16 & 2015-05-04 & 2016-09-28 &  \\
2012-07-17 & 2015-05-10 & 2016-10-02 &  \\
2012-07-21 & 2015-05-18 & 2016-10-10 &  \\
2012-07-26 & 2015-05-23 & 2016-10-14 &  \\
2012-08-01 & 2015-05-26 & 2016-10-18 &  \\
2012-08-05 & 2015-06-05 & 2016-10-22 &  \\
2012-08-10 & 2015-06-09 & 2016-10-26 &  \\
2012-08-11 & 2015-06-12 & 2016-10-30 &  \\
2012-08-13 & 2015-06-15 & 2017-04-19 &  \\
2012-08-13 & 2015-06-18 & 2017-05-03 &  \\
2012-08-14 & 2015-06-21 & 2017-05-17 &  \\
2012-08-15 & 2015-06-24 & 2017-05-31 &  \\
2012-08-19 & 2015-06-27 & 2017-06-14 &  \\
2012-08-20 & 2015-06-29 & 2017-06-28 &  \\
2012-08-24 & 2015-07-02 & 2017-07-12 &  \\
2012-08-30 & 2015-07-06 & 2017-07-26 &  \\
2012-09-07 & 2015-07-09 & 2017-08-08 &. \\
2012-09-14 & 2015-07-12 & 2017-08-23 &  \\
2013-05-14 & 2015-08-10 &	 &	\\
2015-03-17 & 2016-04-17 &	 &	\\
\hline
\end{tabular}
\caption{
Summary of the dates of the existing \emph{Swift}/XRT observations of the globular clusters Terzan~5
and NGC~6388 that were used as input observing campaign in our simulations.
}
\label{tab:observations}
\end{center}
\end{table}

\bibliographystyle{mnras.bst}
\bibliography{./bibliography.bib}

\end{document}